\documentclass[aps,prl,twocolumn,amsmath,amssymb,superscriptaddress,showpacs,10pt]{revtex4-1}

\usepackage[caption=false]{subfig}
\usepackage{graphicx}
\usepackage{mathrsfs} 
\usepackage{bm}
\usepackage[colorlinks]{hyperref}
\usepackage{braket}
\usepackage{soul}
\usepackage{ulem}

\begin{document}

\newcommand{\etal}{\textit{et al}.}
\newcommand{\ms}[1]{\mbox{\scriptsize #1}}
\newcommand{\msb}[1]{\mbox{\scriptsize $\mathbf{#1}$}}
\newcommand{\msi}[1]{\mbox{\scriptsize\textit{#1}}}
\newcommand{\nn}{\nonumber} 
\newcommand{\dg}{^\dagger}
\newcommand{\smallfrac}[2]{\mbox{$\frac{#1}{#2}$}}
\newcommand{\pfpx}[2]{\frac{\partial #1}{\partial #2}}
\newcommand{\dfdx}[2]{\frac{d #1}{d #2}}
\newcommand{\half}{\smallfrac{1}{2}}
\newcommand{\s}{{\mathcal S}}
\newtheorem{theo}{Theorem} \newtheorem{lemma}{Lemma}

\newcommand{\rws}{\color{red}}
\newcommand{\tn}{\color{blue}} 
\definecolor{violet}{rgb}{0.5, 0.0, 0.5}
\definecolor{lukescolor}{rgb}{0.8, 0.33, 0.0}
\definecolor{applegreen}{rgb}{0.55, 0.71, 0.0}
\newcommand{\lrcom}{\color{applegreen}} 
\newcommand{\ak}{\color{violet}} 
\newcommand{\lgcom}{\color{lukescolor}} 
    
\author{T. Noh}
\email{taewan.noh@nist.gov}
\affiliation{Associate of the National Institute of Standards and Technology, Boulder, Colorado 80305, USA}
\affiliation{Department of Physics and Applied Physics, University of Massachusetts, Lowell, MA 01854, USA}
\author{Z. Xiao}
\affiliation{Department of Physics and Applied Physics, University of Massachusetts, Lowell, MA 01854, USA}
\author{K. Cicak}
\affiliation{National Institute of Standards and Technology, 325 Broadway St MS686.05, Boulder, Colorado 80305, USA}
\author{X. Y. Jin}
\affiliation{Associate of the National Institute of Standards and Technology, Boulder, Colorado 80305, USA}
\affiliation{Department of Physics, University of Colorado, Boulder, Colorado 80309, USA}
\author{E. Doucet}
\affiliation{Department of Physics and Applied Physics, University of Massachusetts, Lowell, MA 01854, USA}
\author{J. Teufel}
\affiliation{National Institute of Standards and Technology, 325 Broadway St MS686.05, Boulder, Colorado 80305, USA}
\author{J. Aumentado}
\affiliation{National Institute of Standards and Technology, 325 Broadway St MS686.05, Boulder, Colorado 80305, USA}
\author{L. C. G. Govia}
\affiliation{Quantum Engineering and Computing, Raytheon BBN Technologies, Cambridge, Massachusetts 02138, USA}
\author{L. Ranzani}
\affiliation{Quantum Engineering and Computing, Raytheon BBN Technologies, Cambridge, Massachusetts 02138, USA}
\author{A. Kamal}
\affiliation{Department of Physics and Applied Physics, University of Massachusetts, Lowell, MA 01854, USA}
\author{R. W. Simmonds}
\email{raymond.simmonds@nist.gov}
\affiliation{National Institute of Standards and Technology, 325 Broadway St MS686.05, Boulder, Colorado 80305, USA}

\begin{abstract}
Cavity quantum electrodynamics (QED) with in-situ tunable interactions is important for developing novel systems for quantum simulation and computing.\cite{Rauschenbeutel1999,shane2010,Srinivasan2011,Potts2016, Vaidya2018,Suleymanzade2020} The ability to tune the dispersive shifts of a cavity QED system provides more functionality for performing either quantum measurements or logical manipulations. Here, we couple two transmon qubits to a lumped-element cavity through a shared dc-SQUID.\cite{EandF2011,Lu2017} Our design balances the mutual capacitive and inductive circuit components so that both qubits are highly decoupled from the cavity\cite{Lu2017}, offering protection from decoherence processes\cite{Jed2014,Zhang2017}. We show that by parametrically driving\cite{EandF2011,shane2014,Lu2017} the SQUID with an oscillating flux it is possible to independently tune the interactions between either of the qubits and the cavity dynamically. The strength and detuning of this cavity-QED interaction can be fully controlled through the choice of the parametric pump frequency and amplitude. As a practical demonstration, we perform pulsed parametric dispersive readout of both qubits while statically decoupled from the cavity. The dispersive frequency shifts of the cavity mode follow the expected magnitude and sign based on a simple theory that is supported by a more thorough theoretical investigation\cite{Xiao2021}. This parametric approach creates a new tunable cavity QED framework for developing quantum information systems with various future applications, such as entanglement and error correction via multi-qubit parity readout\cite{Roch2014,Riste2013,Andersen2019}, state and entanglement stabilization\cite{Lu2017,Schwartz2016,Doucet2020}, and parametric logical gates\cite{Reagor2018,Noguchi2018}.
\end{abstract} 

\title{Strong parametric dispersive shifts in a statically decoupled multi-qubit cavity QED system}

\maketitle

Cavity QED has become a backbone for some of the leading quantum information processing systems. In particular, systems based on superconducting circuits, so called circuit QED systems\cite{schoelkopf2004,Koch2007}, provide a promising platform for realizing a practical quantum information processing unit. These systems usually rely on a static coupling between the qubit and cavity.~\cite{StarkShift2005,Houck2008}. In order to overcome the drawbacks in these systems, such as frequency crowding, additional decoherence, and always-on interactions, systems have been developed that use flux tunable elements~\cite{shane2010,Srinivasan2011,shane2014,Jed2014,Zhang2017} to set the coupling strength between a qubit and cavity via a static magnetic field. Tunable couplers have led to improved coherence and fast, high-fidelity two-qubit entangling gates \cite{Chen2014} and qubit measurements have been performed while avoiding decoherence caused by unwanted interactions with the cavity.~\cite{Jed2014,Zhang2017} 

\begin{figure*}[t] 
{\includegraphics[width=1\hsize]{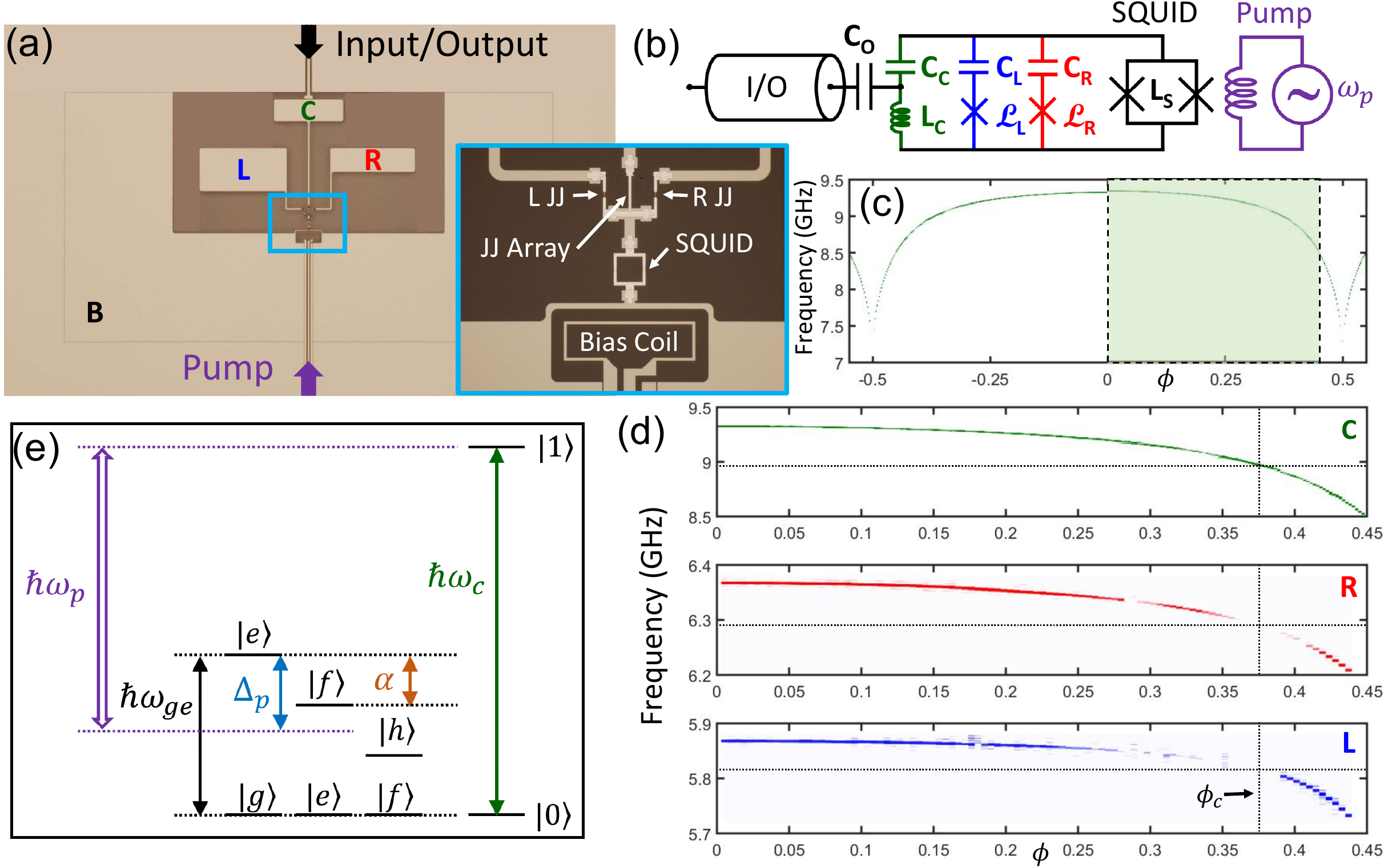}}
\caption{(Color online) (a) Optical micrograph of a tunable circuit QED system. (b) Schematic diagram of the system including the three elements: cavity mode (C), Left transmon (L), and Right transmon (R). The parametric flux drive is applied through a shared dc-SQUID coupler. (c) Periodic cavity spectrum. (d) Frequency of the three resonant modes, C, R, and L as a function of the flux bias $\phi = \Phi/\Phi_0$. (e) Energy scale diagram showing the relationship between defined quantities, see the main text for the details.}
\label{Fig_1} 
\end{figure*}

One difficulty with systems that rely on a tunable static flux is that the frequency of all the coupled elements are usually strongly dependent on the applied flux.\cite{shane2010,Srinivasan2011,shane2014,Jed2014,Zhang2017} Therefore, the tuning of the cavity QED parameters also involves strong manipulation of the frequency of the cavity QED elements. This can be a disadvantage as each element experiences a phase shift when their resonant frequency is adjusted. When performing quantum operations, these shifts must usually be accounted for or corrected. In addition, when many elements share the same frequency space, flux tuning can lead to unwanted frequency collisions and unwanted coupled interactions, complicating the system operation.

Parametric interactions represent an alternative coupling mechanism with many benefits.~\cite{EandF2011,shane2014,Chen2014,Roushan2017} When used in cavity QED-type systems where the parametric pump is resonant with the difference frequency between the qubit and cavity, one can perform rapid swap operations between the coupled systems with rates that are directly proportional to the amplitude of the pump drive.~\cite{EandF2011,shane2014} These type of interactions can be important for cavity photon state manipulation, photon-photon interference, and performing gate operations.~\cite{EandF2011,Nguyen2012,shane2014,Rosenblum2018a}. However, parametric operations within the dispersive regime, where the coupled systems do not exchange energy, can also be very useful. Three-dimensional cavity QED systems have taken advantage of four-wave mixing through a single Josephson junction of a capacitively coupled transmon\cite{Rosenblum2018a} to provide for dynamically adjustable dispersive shifts.~\cite{Rosenblum2018b} Parametric coupling between Qubit-cavity systems can also useful for stabilizing the state of a single transmon qubit~\cite{Lu2017} or the entanglement of multi-qubit systems~\cite{Doucet2020,Brown2021}.

In this work, we describe a two-transmon qubit cavity QED system whereby the interactions with a single cavity mode coupled to an input/output microwave feedline are tuned through the use of a parametric flux drive applied to a shared dc-SQUID coupler. We investigate in detail the dispersive frequency shifts on the cavity mode induced by inserting a single photon in one of the qubits as a function of the parametric pump's drive frequency and amplitude. We find that these shifts can be well described by a simple model that is a parametric generalization of the standard dispersive formulas introduced for circuit QED with a transmon.~\cite{Koch2007} Moreover, we observe experimentally some interesting new features at higher pump amplitudes that can be explained by higher order corrections from the absorption of multiple photons between the transmon's energy levels.~\cite{Xiao2021} In addition, this system also allows parametric interactions between the two qubits, independent of the cavity, for applying parametric entangling gates\cite{Reagor2018,Noguchi2018} or between all three systems for the stabilization of arbitrarily entangled two-qubit states\cite{Schwartz2016,Doucet2020}, as will be described in future work\cite{Jin2021a,Jin2021b,Brown2021}.

With a nearly 0.5~GHz separation between the qubits, the parametric dispersive shifts of the cavity due to either qubit can be independently controlled by selectively applying a specific pump frequency. Both the amplitude and sign of these shifts can be controlled by adjusting the pump amplitude and frequency. Remarkably, we can {\it dynamically} access both positive and negative relative detunings allowing us access to all three distinct dispersive regions~\cite{Koch2007,Xiao2021}, beyond the usual ac Stark effect expected for a two-level atom. Therefore, along with the commonly used negative dispersive shifts, we can also generate large, {\it positive} shifts that hitherto only occurred in the ``straddling regime'' of static circuit QED~\cite{Koch2007}, where the cavity frequency must sit between the first two qubit transition frequencies, $\omega_{ge}$ and $\omega_{ef}$. In this work, by dynamically driving the coupler, a ``{\it parametric} straddling regime'' has been discovered even with the cavity frequency nearly 3~GHz above both qubits. This results from the fact that the {\it parametric} generalization of the qubit-cavity detuning $\Delta_p$, the parametric coupling strength $g_p$, and either transmon anharmonicity $\alpha$ satisfy the necessary straddling regime conditions, making it fully accessible.~\cite{Koch2007,Xiao2021} 

As a practical demonstration of parametrically controlled dispersive shifts, we show that we can pulse ``on'' these shifts in-time in order to measure either qubit, or apply them simultaneously to perform a joint readout of the two-qubit states~\cite{Roch2014,Riste2013,Andersen2019}. One unique design feature of our parametric strategy is that each qubit can be protected~\cite{Srinivasan2011,Zhang2017} from the cavity during its logical operations. Only when it is time to measure the qubit do we turn on the parametric interactions generating large dispersive shifts to enable qubit state readout. To this end, we characterize the residual static dispersive shift between the qubits and the cavity as a function of the flux in the SQUID coupler. Furthermore, by injecting photons into the cavity, we can witness the elimination of additional qubit dephasing even in the presence of photon shot noise in the cavity.~\cite{Zhang2017} These results set the stage for future work. Because these parametric dispersive shifts can be applied simultaneously, with adjustable strength and sign, and add linearly, we can explore the construction of multi-qubit parity measurements, useful for measuring error syndromes and performing quantum error correction.~\cite{Roch2014,Riste2013,Andersen2019}

Our circuit, as shown in Fig.~\ref{Fig_1}(a)--(b), has been constructed to largely cancel static coupling between both qubits and the cavity~\cite{shane2010,shane2014,Lu2017}, so that parametric couplings dominate the interactions (see Supplemental Material (SM)). Each frequency is partially defined by the three top capacitor plates ($C$, $L$, and $R$) for the three lumped-element resonators. The left ($L$) and the right ($R$) transmons each have nominally the same Josephson junction with critical current $I_o\approx 36$~nA that defines the Josephson energy $E_J=I_o\Phi_o/2\pi$. The central cavity ($C$) uses a series array of 7 larger junctions (each with $I_{oC}\approx15I_o$) to form a nearly linear inductance $L_C$. The dc-SQUID coupler, formed by a small loop that includes two Josephson junctions with critical currents $I_{o1}=I_{o2}\approx15I_o$, shares current between all three elements with the large common electrode, or bottom plate $B$. We have purposefully chosen the orientation and distance between the three top capacitor plates in order for their mutual capacitive coupling to cancel the mutual inductive coupling of the SQUID at a particular static flux bias $\phi_c=\Phi_c/\Phi_o$, which we refer to as the ``cancellation flux'', where $\Phi_o = h/2e$ is the magnetic flux quantum. At this flux, the static coupling between all three elements should in-principle be zero, leaving them uncoupled and thus independent. Of course, achieving this condition perfectly is challenging (see SM).

\begin{figure} 
{\includegraphics[width=1\hsize]{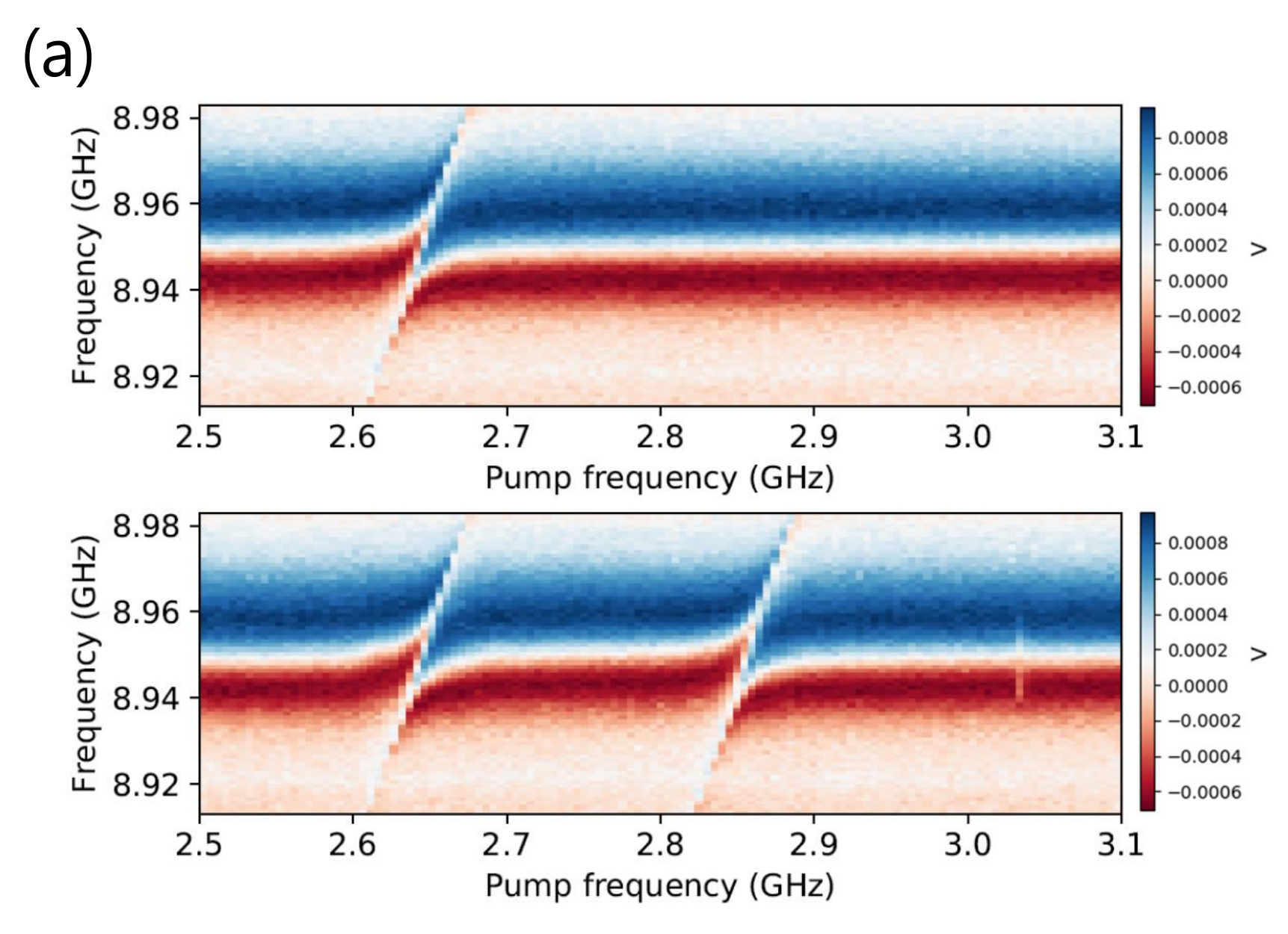}}
{\includegraphics[width=1\hsize]{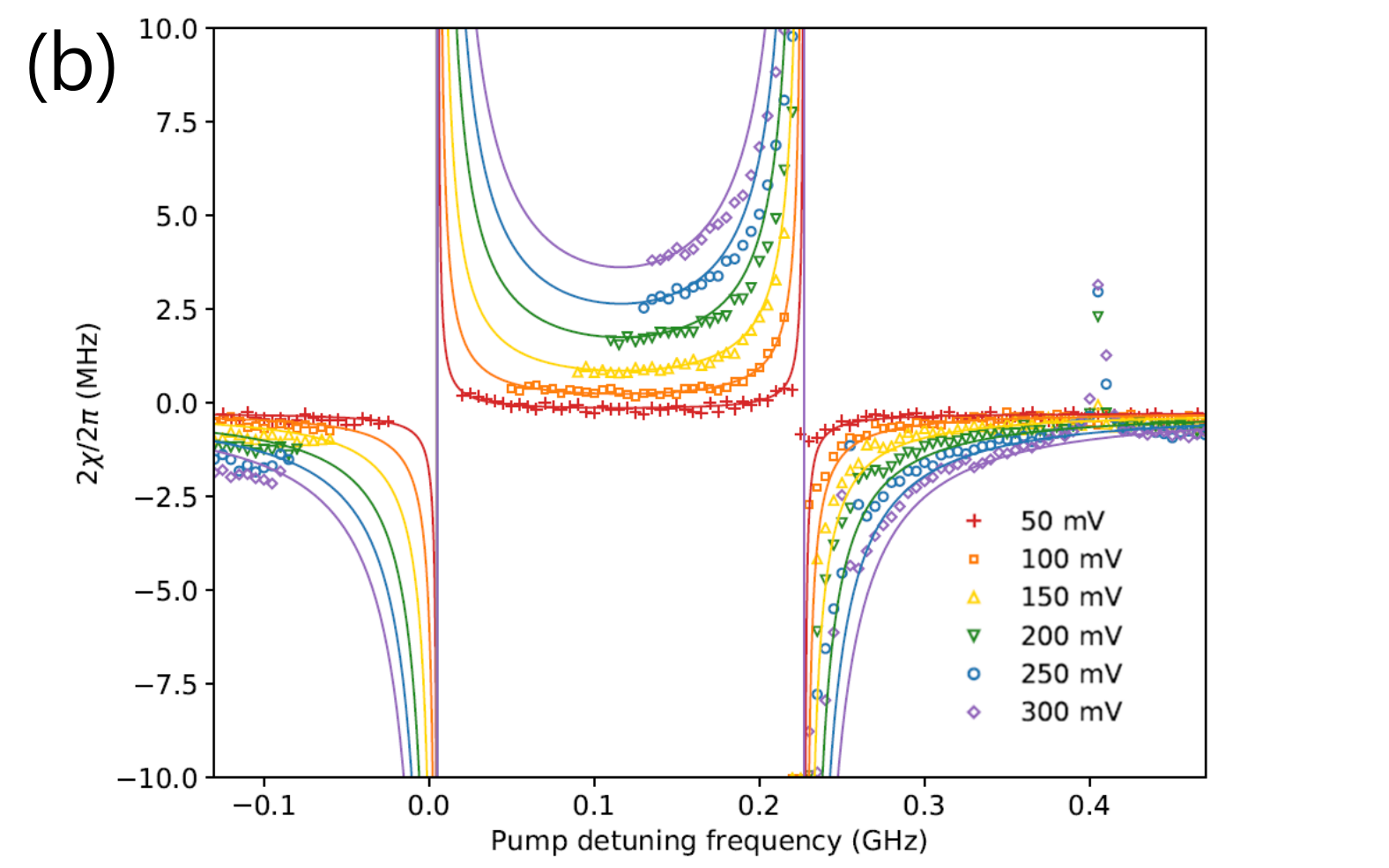}}
{\includegraphics[width=1\hsize]{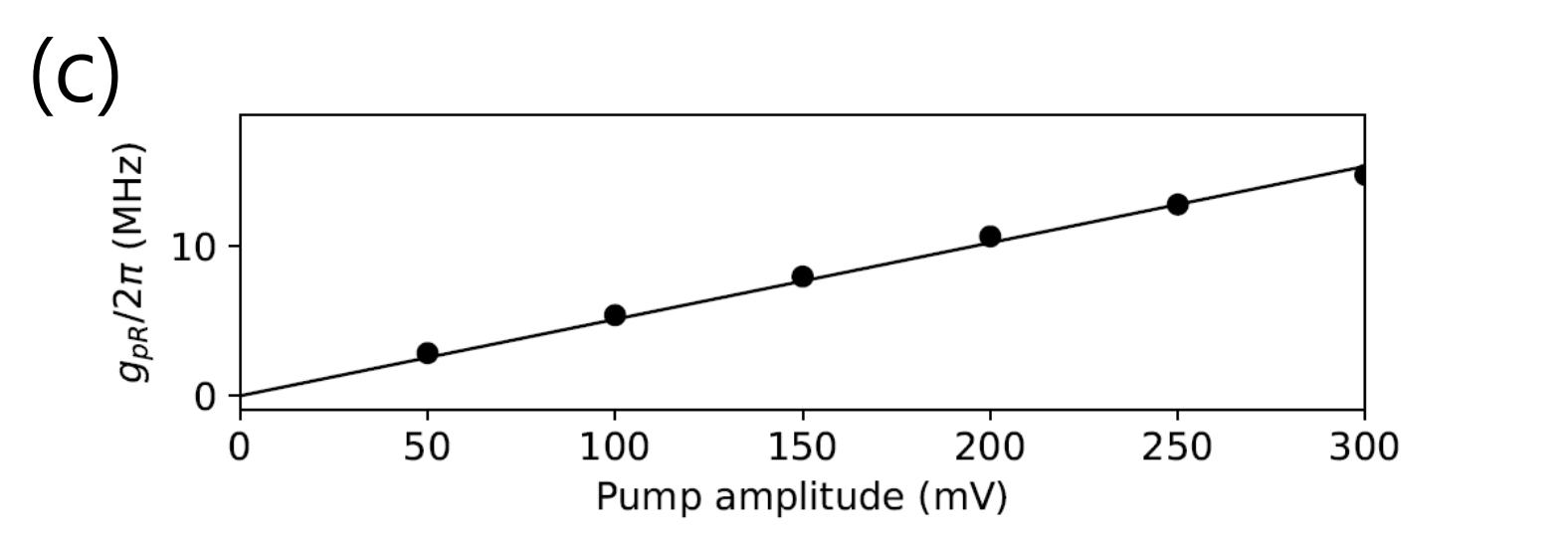}}
\caption{(Color online) (a) A typical full cavity spectrum while sweeping the pump frequency $\omega_p$. The upper and lower panel shows the phase of the cavity signal with the R transmon in state $|g\rangle _R$ and $|e\rangle _R$, respectively. (b) The parametrically induced dispersive shifts as a function of the pump detuning frequency, $\Delta_{pR} = \omega_p - |\Delta_R|$, at various calibrated pump amplitudes: 50 (red), 100 (orange), 150 (yellow), 200 (green), 250 (blue), and 300~mV (purple), respectively. The solid lines are  fits for the data from the theory. See the text for the details. (c) Parametric $g_{pR}$ as a function of pump amplitude.}
\label{Fig_2} 
\end{figure}

Standard circuit QED theory~\cite{schoelkopf2004,Koch2007} describes the interaction between an individual transmon $k=L,R$, with anharmonicity $\alpha_k<0$, and a cavity $C$ as (with $\hbar=1$),
\begin{equation}
    H =  \sum_{k=\{L,R\}} \omega_{k}q_k^\dagger q_k + \frac{\alpha_k}{2} (q_k^\dagger q_k)^2 + \omega_{C}c^\dagger c + H_{Ik},
\label{eq_1}
\end{equation}
with the interaction Hamiltonian is given by,
\begin{equation}
    H_{Ik} = g_{sk}(q_k+q_k^\dagger)(c+c^\dagger),
\label{eq_2}
\end{equation}
where $q_k^\dagger$ ($q_k$) is the creation (annihilation) operator of the $k$th transmon mode, $c^\dagger$ ($c$) corresponds to the same for the cavity mode, $\omega_{k}$ and $\omega_{C}$ are the frequencies of the $k$th transmon and cavity, respectively, and $g_{sk}\equiv g_{sk}(\phi)$ is the SQUID-tunable, total static coupling rate~\cite{shane2010, shane2014,Jed2014,Lu2017} between the $k$th transmon and the cavity. We will define the states $|g\rangle_k$, $|e\rangle_k$ and $|f\rangle_k$ as the first three eigenstates of the $k$th transmon, respectively, while the cavity states are referred to as $|n\rangle$ for photon numbers $n=0,1,2\cdots$. 

In the dispersive regime of cavity QED~\cite{schoelkopf2004,Koch2007,blais2009}, when $g_{sk}\ll|\Delta_k|$ where $\Delta_k=\omega_k-\omega_C$ is the detuning between the $k$th transmon and $g_{sk}$ is the static coupling strength, the interaction Hamiltonian in Eq.~(\ref{eq_2}) simply becomes $H_{Ik}=2\chi_{sk}q_k^\dagger q_k c^\dagger c$. In this regime, the qubit and cavity both experience a total dispersive frequency shift $2\chi_{sk}$ depending on the energy state of the other element of the system. In our case, the flux-tunable {\it{static}} dispersive strength $\chi_{sk}(\phi) = (g_{sk}^2(\phi)/\Delta_k)(\alpha_k/(\alpha_k+\Delta_k))$.~\cite{Koch2007} Here, the separation between the cavity transition frequencies when the $k$th qubit is in $|g\rangle_k$ and $|e\rangle_k$ is $2\chi_{sk}$. At a flux bias far from the cancellation flux, the residual static coupling $g_{sk}$ is predominantly capacitive (when $\phi<\phi_c$) or inductive (when $\phi>\phi_c$). We can easily perform qubit spectroscopy when $\chi_{sk}$ is relatively large, however the spectrum fades away on either side of the cancellation flux as $g_{sk}$ becomes significantly reduced (see Fig.~\ref{Fig_1}(d)).  

The frequencies of all the resonant modes are periodic as a function of $\phi$ (as shown for the cavity in Fig.~\ref{Fig_1}(c)). In Fig.~\ref{Fig_1}(d), we shown all the spectra over nearly a half-a-period in $\phi$.  At $\phi=0$, the maximum qubit frequencies are about 5.9~GHz ($L$) and 6.4~GHz ($R$), while the maximum cavity frequency is about 9.4~GHz. The cavity is coupled to a co-planar waveguide feedline for the input and output of microwave photons leading to a typical total decay rate of $\kappa/2\pi\approx10$~MHz, which is flux dependent (see SM). The transmon anharmonicities are also flux dependent (see SM), with average values of $\alpha_L/2\pi\approx-180$~MHz and $\alpha_R/2\pi\approx-220$~MHz. 

We can access a new, rich set of dynamics by moving beyond the static interactions typical of standard cavity QED systems, by driving the tunable coupler with a sinusoidal flux modulation in addition to a static flux bias,  $\phi(t)=\phi_m\sin(\omega_p t)+\phi_s$. This parametric pump drive then modifies the interaction strength in Eq.~(\ref{eq_2}), leading to the following coupling terms written in an interaction frame w.r.t. the free Hamiltonian~\cite{shane2014},
\begin{equation}
    H_{pk} = -i\sum_{k=\{L,R\}}g_{pk}\left(e^{i\omega_{p}t} - e^{-i\omega_{p}t}\right)(q_k c^\dagger e^{i\Delta_{k}t} + {\rm h.c.}),
\label{eq_3}
\end{equation}
where we have ignored the sum frequency terms for the regime under consideration, i.e. $\omega_p$ near the difference frequency $|\Delta_k|$. In general, the pump frequency can be detuned by $\Delta_{pk}$, $\omega_p=|\Delta_k|+\Delta_{pk}$, in order to sweep between parametrically-induced resonant-type ($\Delta_{pk} \ll g_{pk}$) and dispersive-type ($\Delta_{pk} \gg g_{pk}$) interactions, corresponding to the $|g\rangle _k\rightarrow |e\rangle _k$ transition and the cavity mode. This can been seen by measuring the full spectrum of the cavity mode for a fixed pump amplitude when $\omega_p\approx|\Delta_R|$, as shown in the upper panel of Fig.~\ref{Fig_2}(a). Notice the avoided crossing is clearly visible with a coupling strength, $2g_{pR}/2\pi\approx 10$~MHz. This data was taken when starting from $|g\rangle _R$. If, after a $\pi$-pulse, we start from $|e\rangle _R$ and take the same spectrum in the presence of the pump tone, we can see (lower panel of Fig.~\ref{Fig_2}(a)) another avoided crossing from the $|e\rangle _R\rightarrow |f\rangle _R$ transition becoming resonant with the cavity mode. Notice that this splitting is centered about $\Delta_{pR}=-\alpha_R>0$. This behavior is also seen for higher level transmon transitions~\cite{Xiao2021}, which agrees with Eq.~\ref{eq_4}, and these results have been repeated successfully for the $L$ transmon as well (see SM).

In order to fully characterize the additional dispersive shifts imparted to the cavity by the presence of the parametric pump, we measure the full cavity spectrum for various calibrated pump amplitudes (see SM) while sweeping $\omega_p$. We take two data sets for each pump amplitude, one starting from $|g\rangle _R$ and one starting from $|e\rangle _R$, like the data sets shown in Fig.~\ref{Fig_2}(a)). By comparing the two results, we can extract the full dispersive shifts as shown in Fig.~\ref{Fig_2}(b). For the smallest pump amplitude and largest $\Delta_{pR}$, notice that the dispersive shift is nearly zero. This is because we have chosen $\phi\approx\phi_c$, as discussed below. In general, the parametric dispersive shifts simply add to the static shifts~\cite{Xiao2021} (see SM for data at more bias points).

\begin{figure}[t]
{\includegraphics[width=1\hsize]{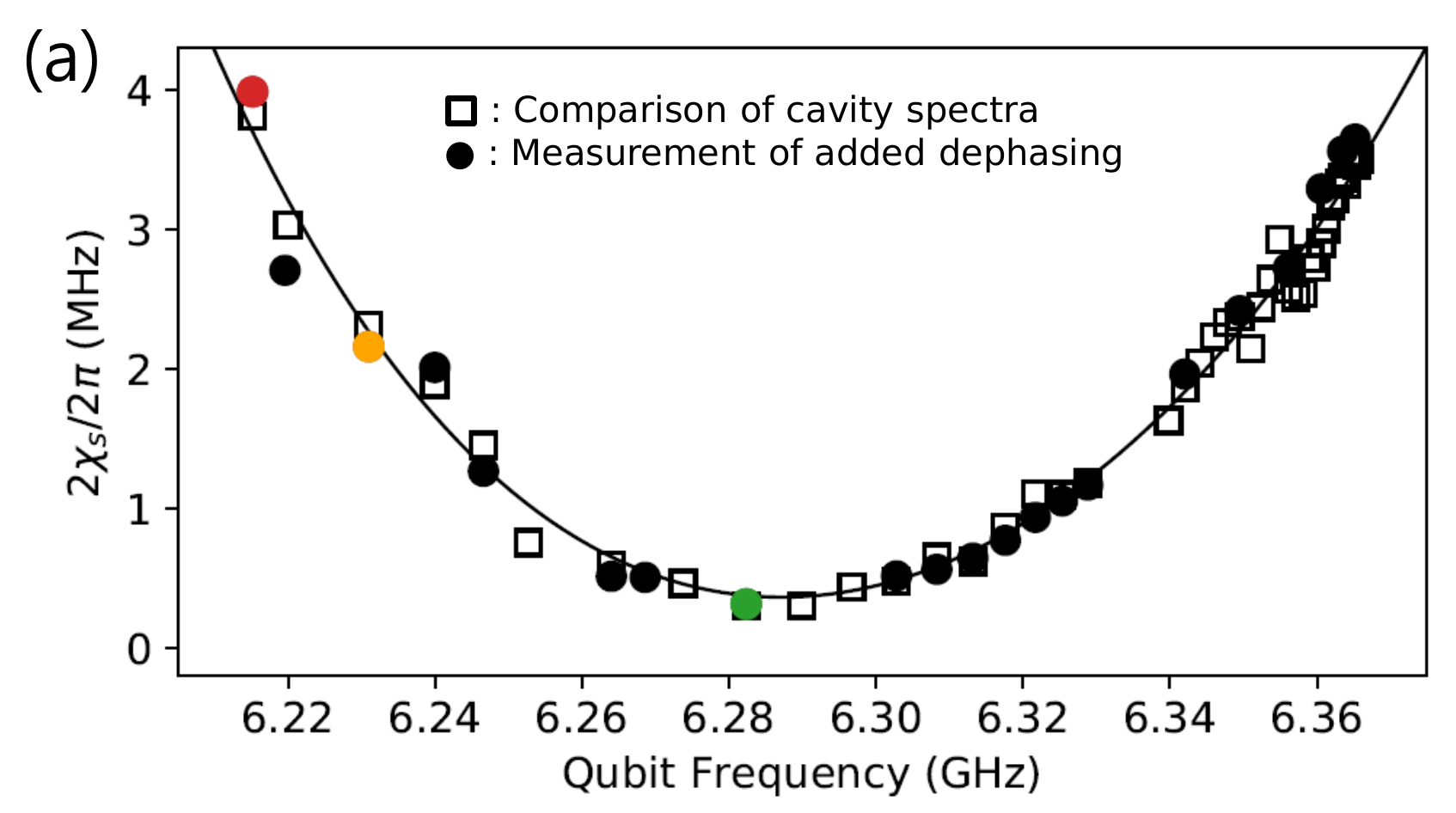}}
{\includegraphics[width=1\hsize]{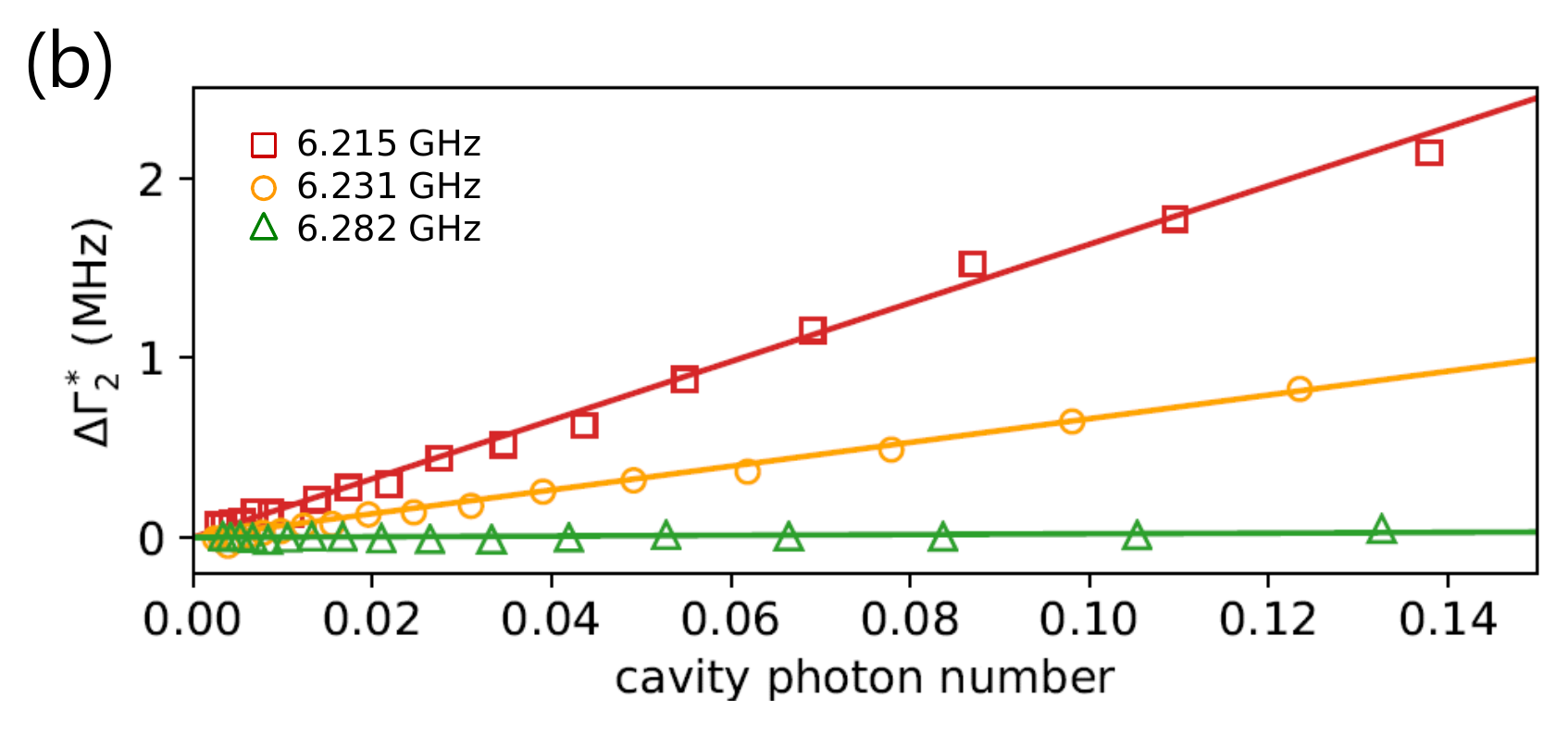}}
\caption{(Color online) (a) The static dispersive shift $2\chi_s/2\pi$ as a function of the R transmon frequency. For red open squares, the dispersive shift was measured by comparing the cavity response with the R transmon in either state $|g\rangle _R$ or $|e\rangle _R$. For blue filled circles, the dispersive shift was extracted by measuring additional dephasing $\Gamma_{\phi R}$ while driving the cavity to a weak coherent state, see the main text for the details. The solid line represents a polynomial fit to the data. (b) Additional dephasing as a function of the average photon number $\bar{n}$ at three different flux biases where the R tranmon frequencies are 6.215 (red square), 6.231 (yellow circle), and 6.282~GHz (green triangle), respectively. Solid lines are a linear fit for the data.}
\label{Fig_3} 
\end{figure}

In order to gain physical intuition, the parametric interactions can be analyzed in a rotating frame defined w.r.t. the pump frequency $\omega_{p}$. For an arbitrary pump detuning $\Delta_{pk}$, this necessitates including contributions from both positive and negative frequency components, as well as those corresponding to the sum and difference frequencies.~\cite{Xiao2021} However, for sufficiently small $\Delta_{pk}$, in this rotating frame, the form of Eq.~(\ref{eq_3}) becomes analogous to the static dispersive case with the identification $g_{sk}\rightarrow g_{pk}$ and $\Delta_k\rightarrow \Delta_{pk}$.~\cite{Koch2007} We can even extend this picture to include higher transmon levels (see SM),

\begin{equation}
    \chi_{pk} = \frac{g_{pk}^2}{\Delta_{pk}}\left[\frac{\alpha_k}{(\alpha_k+\Delta_{pk})}-\frac{\Delta_{pk}}{(2\alpha_k+\Delta_{pk})}-\frac{\Delta_{pk}}{(3\alpha_k+\Delta_{pk})}\cdots\right],
\label{eq_4}
\end{equation}

This simplified expression describes the data quite well and the extracted values for $g_{pR}$ are plotted in Fig.~\ref{Fig_2}(c), which clearly shows a linear dependence on the pump amplitude. Moreover, at a fixed amplitude, $g_{pR}\propto\sqrt{(d\omega_R/d\phi)(d\omega_C/d\phi)}$ agrees with predictions~\cite{EandF2011,shane2014} based on the slope of the qubit and cavity at the cancellation flux $\phi_c$ (see SM). At larger powers, however, we find a very interesting peak in the data that occurs at $\Delta_{pR}=-2\alpha_R$. This corresponds to a pump-mediated two-photon transition between the first- and third-excited states of the transmon~\cite{Xiao2021}. Such features are described by quartic-order contributions to dispersive shift $\chi_{pk}$ and exhibit an {\it even} lineshape in pump detuning, thus providing a means to distinguish them from the lower-order quadratic features that lead to an {\it odd} lineshape (see Fig.~\ref{Fig_2}(b)). 

As a simple demonstration of the usefulness of programmable dispersive shifts in this new parametric circuit-QED framework, we next describe qubit measurements near the cancellation flux. One drawback when working with static cavity QED systems is that a persistent dispersive interaction can lead to two decoherence processes, enhanced energy relaxation~\cite{Houck2008}, reducing the qubit relaxation time $T_1$, and additional dephasing~\cite{StarkShift2005}, reducing the qubit dephasing time $T_2$. Although the qubit and the cavity are detuned from each other, the qubit’s decay rate $1/T_{1}$ is enhanced by an additional Purcell factor of $(g/\Delta)^2\kappa$ due to radiative decay through the coupled cavity mode. Whereas qubit dephasing occurs because the dispersive interaction also modifies the qubit’s frequency depending on the photon occupancy within the cavity. The additional dephasing is given by the rate $\Gamma_n=8\bar{n}\kappa\chi_s^2/(\kappa^2 + 4\chi_s^2)$, where $\bar{n}$ is the average number of (coherent state) photons in the cavity.~\cite{StarkShift2005,Gambetta2006} These decoherence mechanisms can be reduced in circuit QED systems by using  Purcell filters~\cite{Reed2010} (increases $T_1$ only) or eliminating the total qubit-cavity coupling~\cite{shane2010,Srinivasan2011,shane2014,Zhang2017} (increases both $T_1$ and $T_2$). One difficulty with the second approach is that once $g_{sk}\approx 0$, it is not possible to measure the qubit dispersively, unless the qubit has hidden degrees of freedom still coupled to the cavity~\cite{Zhang2017} or you rapidly tune-up $g_{sk}(\phi)$ through a flux shift pulse in time for the measurement~\cite{Jed2014}. 

Alternatively, our circuit allows us to minimize $g_{sk}$ by setting $\phi=\phi_c$, while still enabling dispersive qubit readout by applying a parametric pump tone with a given amplitude and frequency $\omega_p$, as described above (see Fig.~\ref{Fig_2}). Ideally, when $\chi_s\propto g_{sk}^2\rightarrow 0$, both qubits are completely decoupled statically from the cavity. In Fig.~\ref{Fig_3}(a), we show the static dispersive shift $\chi_{sR}$ (and $g_{sR}$, see SM) as a function of the R transmon frequency. For the open square, we directly measured the dispersive shift by comparing the cavity response with the R transmon in either state $|g\rangle _R$ or $|e\rangle _R$. The solid line comes from a polynomial fit that matches our predictions based on a circuit model (see SM). In Fig.~\ref{Fig_3}(b), we also measured additional dephasing $\Gamma_{\phi R}$ of the R transmon while driving the cavity with a weak coherent state with an average photon number $\bar{n}$ for each flux bias.~\cite{StarkShift2005,Zhang2017} Crucially, the measurement of the R transmon near the cancellation flux required using a parametric tone to perform the dispersive readout. We see that $\Gamma_{\phi R}$ scales linearly with $\bar{n}$ as expected~\cite{StarkShift2005,Gambetta2006} and the slope significantly diminishes as the bias approaches the cancellation flux. From this data, and knowing $\kappa$, we can extract $\chi_{sR}$ (and $g_{sR}$) for each bias (see SM).  This result is plotted as filled circles in Fig.~\ref{Fig_3}(a) and agrees well with the direct measurements. These results were also performed on the L transmon with similar results (see SM). The minimum value for $\chi_{sR}/2\pi$ ($\chi_{sL}/2\pi$) was $300$~kHz ($150$~kHz), which was limited by the nonlinearity of the SQUID coupler itself (see SM).

\begin{figure}[t] 
{\includegraphics[width=1\hsize]{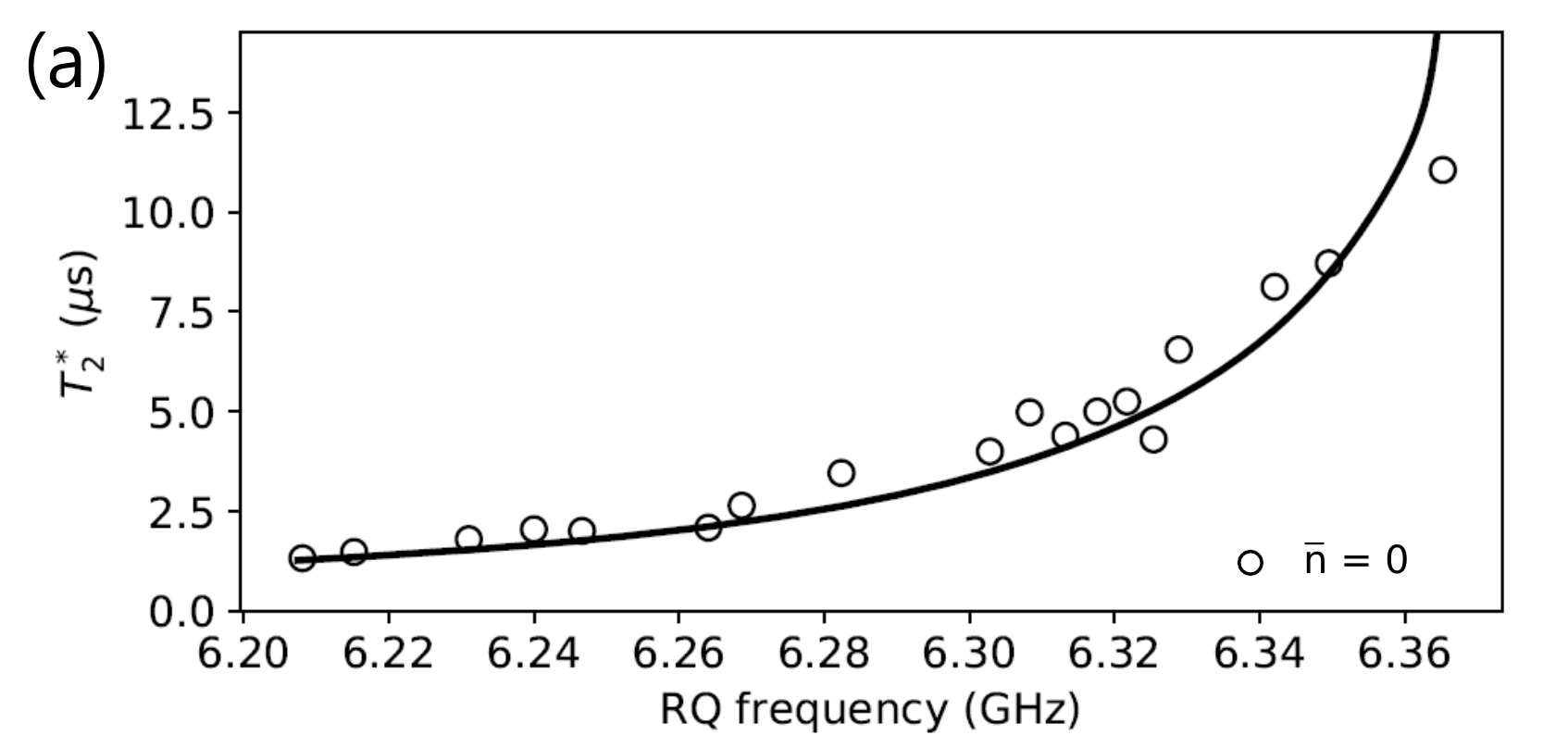}}
{\includegraphics[width=1\hsize]{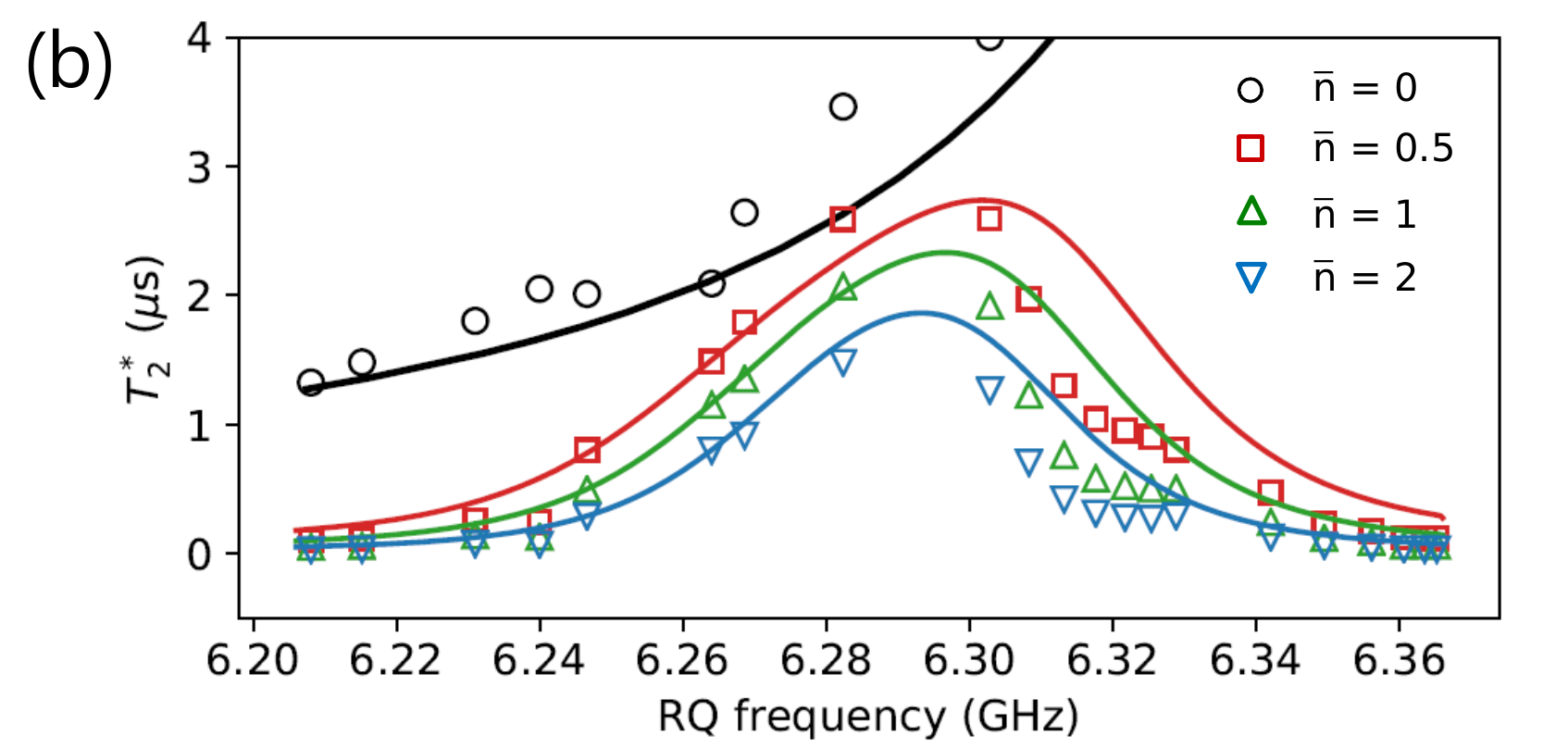}}
\caption{(Color online) (a) The phase coherence time $T^*_2$ as a function of the R transmon frequency when $\bar{n} \approx 0$. The solid black line is a fit based on the dominant dephasing coming from flux noise. (b) Comparison of $T^*_2$ in the presence of coherent cavity photons with $\bar{n} =$ 0.5 (square), 1 (up-triangle), and 2 (down-triangle). The solid lines are predictions based the measurement, see the main text for the details.}
\label{Fig_4} 
\end{figure}

Finally, we can demonstrate that the qubits are protected from photon shot noise in the cavity by comparing the additional dephasing just discussed ($\bar{n}\neq 0$) with the background dephasing when $\bar{n}\approx 0$.~\cite{Zhang2017} In Fig.~\ref{Fig_4}(a), we plot the phase coherence time $T^*_2$ of the R~qubit obtained by measuring Ramsey oscillations as a function of its frequency. The solid line is a fit to the data assuming that the dominant source of background dephasing is coming from flux noise in the coupler, which is proportional to the slope $d{\omega_R(\phi)}/d\phi$ of the modulation curve of R transmon frequency with flux (see Fig.~\ref{Fig_1}). In Fig.~\ref{Fig_4}(b), we compare this result with $T^*_2$ of the R transmon in the presence of photon shot noise for $\bar{n}=$0, 0.5, 1, and 2. Here, the solid lines are predictions based on the directly measured values for $\chi_{sR}$ as a function of flux and the average value of $\kappa$ over whole range of flux considered in the measurement. As we can see, the R transmon is protected from shot noise in the cavity near $\phi=\phi_c$, where the dephasing is dominated by flux noise in the coupler. Similar results were found for the L transmon as well (See SM).  

In conclusion, we have developed a unique tunable cavity QED system that allows for in-situ dynamic control over the interaction between a qubit and a cavity by applying a parametric tone. We demonstrated that this can be accomplished with a two-qubit system, that can also allow for a fully adjustable joint two-qubit measurement. We operated our two-transmon circuit QED system in the dispersive regime and explored the dynamic control of parametrically induced dispersive cavity shifts from either transmon. By varying the sign and size of these shifts with the amplitude and frequency of the parametric pump, we have unlocked a ``{\it parametric} straddling regime'' and verified a simple theoretical model that generalizes this behavior in the rotating frame. As a practical application, we pulsed the parametric pump simultaneously with a readout tone, and with a proper choice for the amplitude and frequency, we performed dispersive qubit readout on both transmons. Due to our unique circuit design that minimizes static coupling to the cavity, the transmons were protected during local logical operations from photon shot noise in the cavity. We believe that this new tunable cavity QED framework will open up a novel paradigm for controlling light-matter interactions by providing a unique functionality along with qualitatively new features that are not supported by static circuit-QED setups.

This work was partially performed under the following financial assistance award 70NANB18H006 from U.S. Department of Commerce, National Institute of Standards and Technology. T.N., Z.X.,E.D.,L.R.,L.G. and A.K. received support from the Department of Energy under grant DE-SC0019461.

%


\begin{thebibliography}{34}%
\makeatletter
\providecommand \@ifxundefined [1]{%
 \@ifx{#1\undefined}
}%
\providecommand \@ifnum [1]{%
 \ifnum #1\expandafter \@firstoftwo
 \else \expandafter \@secondoftwo
 \fi
}%
\providecommand \@ifx [1]{%
 \ifx #1\expandafter \@firstoftwo
 \else \expandafter \@secondoftwo
 \fi
}%
\providecommand \natexlab [1]{#1}%
\providecommand \enquote  [1]{``#1''}%
\providecommand \bibnamefont  [1]{#1}%
\providecommand \bibfnamefont [1]{#1}%
\providecommand \citenamefont [1]{#1}%
\providecommand \href@noop [0]{\@secondoftwo}%
\providecommand \href [0]{\begingroup \@sanitize@url \@href}%
\providecommand \@href[1]{\@@startlink{#1}\@@href}%
\providecommand \@@href[1]{\endgroup#1\@@endlink}%
\providecommand \@sanitize@url [0]{\catcode `\\12\catcode `\$12\catcode
  `\&12\catcode `\#12\catcode `\^12\catcode `\_12\catcode `\%12\relax}%
\providecommand \@@startlink[1]{}%
\providecommand \@@endlink[0]{}%
\providecommand \url  [0]{\begingroup\@sanitize@url \@url }%
\providecommand \@url [1]{\endgroup\@href {#1}{\urlprefix }}%
\providecommand \urlprefix  [0]{URL }%
\providecommand \Eprint [0]{\href }%
\providecommand \doibase [0]{http://dx.doi.org/}%
\providecommand \selectlanguage [0]{\@gobble}%
\providecommand \bibinfo  [0]{\@secondoftwo}%
\providecommand \bibfield  [0]{\@secondoftwo}%
\providecommand \translation [1]{[#1]}%
\providecommand \BibitemOpen [0]{}%
\providecommand \bibitemStop [0]{}%
\providecommand \bibitemNoStop [0]{.\EOS\space}%
\providecommand \EOS [0]{\spacefactor3000\relax}%
\providecommand \BibitemShut  [1]{\csname bibitem#1\endcsname}%
\let\auto@bib@innerbib\@empty
\bibitem [{\citenamefont {Rauschenbeutel}\ \emph {et~al.}(1999)\citenamefont
  {Rauschenbeutel}, \citenamefont {Nogues}, \citenamefont {Osnaghi},
  \citenamefont {Bertet}, \citenamefont {Brune}, \citenamefont {Raimond},\ and\
  \citenamefont {Haroche}}]{Rauschenbeutel1999}%
  \BibitemOpen
  \bibfield  {author} {\bibinfo {author} {\bibfnamefont {A.}~\bibnamefont
  {Rauschenbeutel}}, \bibinfo {author} {\bibfnamefont {A.~G.}\ \bibnamefont
  {Nogues}}, \bibinfo {author} {\bibfnamefont {S.}~\bibnamefont {Osnaghi}},
  \bibinfo {author} {\bibfnamefont {P.}~\bibnamefont {Bertet}}, \bibinfo
  {author} {\bibfnamefont {M.}~\bibnamefont {Brune}}, \bibinfo {author}
  {\bibfnamefont {J.~M.}\ \bibnamefont {Raimond}}, \ and\ \bibinfo {author}
  {\bibfnamefont {S.}~\bibnamefont {Haroche}},\ }\href {\doibase
  https://doi.org/10.1103/PhysRevLett.83.5166} {\bibfield  {journal} {\bibinfo
  {journal} {Phys. Rev. Lett.}\ }\textbf {\bibinfo {volume} {83}},\ \bibinfo
  {pages} {5166} (\bibinfo {year} {1999})}\BibitemShut {NoStop}%
\bibitem [{\citenamefont {Allman}\ \emph {et~al.}(2010)\citenamefont {Allman},
  \citenamefont {Altomare}, \citenamefont {Whittaker}, \citenamefont {Cicak},
  \citenamefont {Li}, \citenamefont {Sirois}, \citenamefont {Strong},
  \citenamefont {Teufel},\ and\ \citenamefont {Simmonds}}]{shane2010}%
  \BibitemOpen
  \bibfield  {author} {\bibinfo {author} {\bibfnamefont {M.}~\bibnamefont
  {Allman}}, \bibinfo {author} {\bibfnamefont {F.}~\bibnamefont {Altomare}},
  \bibinfo {author} {\bibfnamefont {J.}~\bibnamefont {Whittaker}}, \bibinfo
  {author} {\bibfnamefont {K.}~\bibnamefont {Cicak}}, \bibinfo {author}
  {\bibfnamefont {D.}~\bibnamefont {Li}}, \bibinfo {author} {\bibfnamefont
  {A.}~\bibnamefont {Sirois}}, \bibinfo {author} {\bibfnamefont
  {J.}~\bibnamefont {Strong}}, \bibinfo {author} {\bibfnamefont
  {J.}~\bibnamefont {Teufel}}, \ and\ \bibinfo {author} {\bibfnamefont
  {R.}~\bibnamefont {Simmonds}},\ }\href@noop {} {\bibfield  {journal}
  {\bibinfo  {journal} {Phys. Rev. Lett.}\ }\textbf {\bibinfo {volume} {104}},\
  \bibinfo {pages} {177004} (\bibinfo {year} {2010})}\BibitemShut {NoStop}%
\bibitem [{\citenamefont {Srinivasan}\ \emph {et~al.}(2011)\citenamefont
  {Srinivasan}, \citenamefont {Hoffman}, \citenamefont {Gambetta},\ and\
  \citenamefont {Houck}}]{Srinivasan2011}%
  \BibitemOpen
  \bibfield  {author} {\bibinfo {author} {\bibfnamefont {S.~J.}\ \bibnamefont
  {Srinivasan}}, \bibinfo {author} {\bibfnamefont {A.~J.}\ \bibnamefont
  {Hoffman}}, \bibinfo {author} {\bibfnamefont {J.~M.}\ \bibnamefont
  {Gambetta}}, \ and\ \bibinfo {author} {\bibfnamefont {A.~A.}\ \bibnamefont
  {Houck}},\ }\href@noop {} {\bibfield  {journal} {\bibinfo  {journal} {Phys.
  Rev. Lett.}\ }\textbf {\bibinfo {volume} {106}},\ \bibinfo {pages} {083601}
  (\bibinfo {year} {2011})}\BibitemShut {NoStop}%
\bibitem [{\citenamefont {Potts}\ \emph {et~al.}(2016)\citenamefont {Potts},
  \citenamefont {Melnyk}, \citenamefont {Ramp}, \citenamefont {Bitarafan},
  \citenamefont {Vick}, \citenamefont {LeBlanc}, \citenamefont {Davis},\ and\
  \citenamefont {DeCorby}}]{Potts2016}%
  \BibitemOpen
  \bibfield  {author} {\bibinfo {author} {\bibfnamefont {C.~A.}\ \bibnamefont
  {Potts}}, \bibinfo {author} {\bibfnamefont {A.}~\bibnamefont {Melnyk}},
  \bibinfo {author} {\bibfnamefont {H.}~\bibnamefont {Ramp}}, \bibinfo {author}
  {\bibfnamefont {M.~H.}\ \bibnamefont {Bitarafan}}, \bibinfo {author}
  {\bibfnamefont {D.}~\bibnamefont {Vick}}, \bibinfo {author} {\bibfnamefont
  {L.~J.}\ \bibnamefont {LeBlanc}}, \bibinfo {author} {\bibfnamefont {J.~P.}\
  \bibnamefont {Davis}}, \ and\ \bibinfo {author} {\bibfnamefont {R.~G.}\
  \bibnamefont {DeCorby}},\ }\href@noop {} {\bibfield  {journal} {\bibinfo
  {journal} {Appl. Phys. Lett.}\ }\textbf {\bibinfo {volume} {108}},\ \bibinfo
  {pages} {041103} (\bibinfo {year} {2016})}\BibitemShut {NoStop}%
\bibitem [{\citenamefont {Vaidya}\ \emph {et~al.}(2018)\citenamefont {Vaidya},
  \citenamefont {Guo}, \citenamefont {Kroeze}, \citenamefont {Ballantine},
  \citenamefont {Koll\'{a}r}, \citenamefont {Keeling},\ and\ \citenamefont
  {Lev}}]{Vaidya2018}%
  \BibitemOpen
  \bibfield  {author} {\bibinfo {author} {\bibfnamefont {V.~D.}\ \bibnamefont
  {Vaidya}}, \bibinfo {author} {\bibfnamefont {Y.}~\bibnamefont {Guo}},
  \bibinfo {author} {\bibfnamefont {R.~M.}\ \bibnamefont {Kroeze}}, \bibinfo
  {author} {\bibfnamefont {K.~E.}\ \bibnamefont {Ballantine}}, \bibinfo
  {author} {\bibfnamefont {A.~J.}\ \bibnamefont {Koll\'{a}r}}, \bibinfo
  {author} {\bibfnamefont {J.}~\bibnamefont {Keeling}}, \ and\ \bibinfo
  {author} {\bibfnamefont {B.~L.}\ \bibnamefont {Lev}},\ }\href@noop {}
  {\bibfield  {journal} {\bibinfo  {journal} {Phys. Rev. X}\ }\textbf {\bibinfo
  {volume} {8}},\ \bibinfo {pages} {011002} (\bibinfo {year}
  {2018})}\BibitemShut {NoStop}%
\bibitem [{\citenamefont {Suleymanzade}\ \emph {et~al.}(2020)\citenamefont
  {Suleymanzade}, \citenamefont {Anferov}, \citenamefont {Stone}, \citenamefont
  {Naik}, \citenamefont {Oriani}, \citenamefont {Simon},\ and\ \citenamefont
  {Schuster}}]{Suleymanzade2020}%
  \BibitemOpen
  \bibfield  {author} {\bibinfo {author} {\bibfnamefont {A.}~\bibnamefont
  {Suleymanzade}}, \bibinfo {author} {\bibfnamefont {A.}~\bibnamefont
  {Anferov}}, \bibinfo {author} {\bibfnamefont {M.}~\bibnamefont {Stone}},
  \bibinfo {author} {\bibfnamefont {R.~K.}\ \bibnamefont {Naik}}, \bibinfo
  {author} {\bibfnamefont {A.}~\bibnamefont {Oriani}}, \bibinfo {author}
  {\bibfnamefont {J.}~\bibnamefont {Simon}}, \ and\ \bibinfo {author}
  {\bibfnamefont {D.}~\bibnamefont {Schuster}},\ }\href@noop {} {\bibfield
  {journal} {\bibinfo  {journal} {Appl. Phys. Lett.}\ }\textbf {\bibinfo
  {volume} {116}},\ \bibinfo {pages} {104001} (\bibinfo {year}
  {2020})}\BibitemShut {NoStop}%
\bibitem [{\citenamefont {Zakka-Bajjani}\ \emph {et~al.}(2011)\citenamefont
  {Zakka-Bajjani}, \citenamefont {Nguyen}, \citenamefont {Lee}, \citenamefont
  {Vale}, \citenamefont {Simmonds},\ and\ \citenamefont
  {Aumentado}}]{EandF2011}%
  \BibitemOpen
  \bibfield  {author} {\bibinfo {author} {\bibfnamefont {E.}~\bibnamefont
  {Zakka-Bajjani}}, \bibinfo {author} {\bibfnamefont {F.}~\bibnamefont
  {Nguyen}}, \bibinfo {author} {\bibfnamefont {M.}~\bibnamefont {Lee}},
  \bibinfo {author} {\bibfnamefont {L.~R.}\ \bibnamefont {Vale}}, \bibinfo
  {author} {\bibfnamefont {R.~W.}\ \bibnamefont {Simmonds}}, \ and\ \bibinfo
  {author} {\bibfnamefont {J.}~\bibnamefont {Aumentado}},\ }\href@noop {}
  {\bibfield  {journal} {\bibinfo  {journal} {Nature Physics}\ }\textbf
  {\bibinfo {volume} {7}},\ \bibinfo {pages} {599} (\bibinfo {year}
  {2011})}\BibitemShut {NoStop}%
\bibitem [{\citenamefont {Lu}\ \emph {et~al.}(2017)\citenamefont {Lu},
  \citenamefont {Chakram}, \citenamefont {Leung}, \citenamefont {Earnest},
  \citenamefont {Naik}, \citenamefont {Huang}, \citenamefont {Groszkowski},
  \citenamefont {Kapit}, \citenamefont {Koch},\ and\ \citenamefont
  {Schuster}}]{Lu2017}%
  \BibitemOpen
  \bibfield  {author} {\bibinfo {author} {\bibfnamefont {Y.}~\bibnamefont
  {Lu}}, \bibinfo {author} {\bibfnamefont {S.}~\bibnamefont {Chakram}},
  \bibinfo {author} {\bibfnamefont {N.}~\bibnamefont {Leung}}, \bibinfo
  {author} {\bibfnamefont {N.}~\bibnamefont {Earnest}}, \bibinfo {author}
  {\bibfnamefont {R.~K.}\ \bibnamefont {Naik}}, \bibinfo {author}
  {\bibfnamefont {Z.}~\bibnamefont {Huang}}, \bibinfo {author} {\bibfnamefont
  {P.}~\bibnamefont {Groszkowski}}, \bibinfo {author} {\bibfnamefont
  {E.}~\bibnamefont {Kapit}}, \bibinfo {author} {\bibfnamefont
  {J.}~\bibnamefont {Koch}}, \ and\ \bibinfo {author} {\bibfnamefont {D.~I.}\
  \bibnamefont {Schuster}},\ }\href@noop {} {\bibfield  {journal} {\bibinfo
  {journal} {Phys. Rev. Lett.}\ }\textbf {\bibinfo {volume} {119}},\ \bibinfo
  {pages} {150502} (\bibinfo {year} {2017})}\BibitemShut {NoStop}%
\bibitem [{\citenamefont {Whittaker}\ \emph {et~al.}(2014)\citenamefont
  {Whittaker}, \citenamefont {da~Silva}, \citenamefont {Allman}, \citenamefont
  {Lecocq}, \citenamefont {Cicak}, \citenamefont {Sirois}, \citenamefont
  {Teufel}, \citenamefont {Aumentado},\ and\ \citenamefont
  {Simmonds}}]{Jed2014}%
  \BibitemOpen
  \bibfield  {author} {\bibinfo {author} {\bibfnamefont {J.~D.}\ \bibnamefont
  {Whittaker}}, \bibinfo {author} {\bibfnamefont {F.~C.~S.}\ \bibnamefont
  {da~Silva}}, \bibinfo {author} {\bibfnamefont {M.~S.}\ \bibnamefont
  {Allman}}, \bibinfo {author} {\bibfnamefont {F.}~\bibnamefont {Lecocq}},
  \bibinfo {author} {\bibfnamefont {K.}~\bibnamefont {Cicak}}, \bibinfo
  {author} {\bibfnamefont {A.~J.}\ \bibnamefont {Sirois}}, \bibinfo {author}
  {\bibfnamefont {J.~D.}\ \bibnamefont {Teufel}}, \bibinfo {author}
  {\bibfnamefont {J.}~\bibnamefont {Aumentado}}, \ and\ \bibinfo {author}
  {\bibfnamefont {R.~W.}\ \bibnamefont {Simmonds}},\ }\href@noop {} {\bibfield
  {journal} {\bibinfo  {journal} {Phys. Rev. B}\ }\textbf {\bibinfo {volume}
  {90}},\ \bibinfo {pages} {024513} (\bibinfo {year} {2014})}\BibitemShut
  {NoStop}%
\bibitem [{\citenamefont {Zhang}\ \emph {et~al.}(2017)\citenamefont {Zhang},
  \citenamefont {Liu}, \citenamefont {Raftery},\ and\ \citenamefont
  {Houck}}]{Zhang2017}%
  \BibitemOpen
  \bibfield  {author} {\bibinfo {author} {\bibfnamefont {G.}~\bibnamefont
  {Zhang}}, \bibinfo {author} {\bibfnamefont {Y.}~\bibnamefont {Liu}}, \bibinfo
  {author} {\bibfnamefont {J.~J.}\ \bibnamefont {Raftery}}, \ and\ \bibinfo
  {author} {\bibfnamefont {A.~A.}\ \bibnamefont {Houck}},\ }\href {\doibase
  https://doi.org/10.1038/s41534-016-0002-2} {\bibfield  {journal} {\bibinfo
  {journal} {npj Quantum Information}\ }\textbf {\bibinfo {volume} {3}},\
  \bibinfo {pages} {1} (\bibinfo {year} {2017})}\BibitemShut {NoStop}%
\bibitem [{\citenamefont {Allman}\ \emph {et~al.}(2014)\citenamefont {Allman},
  \citenamefont {Whittaker}, \citenamefont {Castellanos-Beltran}, \citenamefont
  {Cicak}, \citenamefont {da~Silva}, \citenamefont {DeFeo}, \citenamefont
  {Lecocq}, \citenamefont {Sirois}, \citenamefont {Teufel}, \citenamefont
  {Aumentado},\ and\ \citenamefont {Simmonds}}]{shane2014}%
  \BibitemOpen
  \bibfield  {author} {\bibinfo {author} {\bibfnamefont {M.}~\bibnamefont
  {Allman}}, \bibinfo {author} {\bibfnamefont {J.}~\bibnamefont {Whittaker}},
  \bibinfo {author} {\bibfnamefont {M.}~\bibnamefont {Castellanos-Beltran}},
  \bibinfo {author} {\bibfnamefont {K.}~\bibnamefont {Cicak}}, \bibinfo
  {author} {\bibfnamefont {F.}~\bibnamefont {da~Silva}}, \bibinfo {author}
  {\bibfnamefont {M.}~\bibnamefont {DeFeo}}, \bibinfo {author} {\bibfnamefont
  {F.}~\bibnamefont {Lecocq}}, \bibinfo {author} {\bibfnamefont
  {A.}~\bibnamefont {Sirois}}, \bibinfo {author} {\bibfnamefont
  {J.}~\bibnamefont {Teufel}}, \bibinfo {author} {\bibfnamefont
  {J.}~\bibnamefont {Aumentado}}, \ and\ \bibinfo {author} {\bibfnamefont
  {R.}~\bibnamefont {Simmonds}},\ }\href@noop {} {\bibfield  {journal}
  {\bibinfo  {journal} {Phys. Rev. Lett.}\ }\textbf {\bibinfo {volume} {112}},\
  \bibinfo {pages} {123601} (\bibinfo {year} {2014})}\BibitemShut {NoStop}%
\bibitem [{\citenamefont {Xiao}\ \emph {et~al.}(2021)\citenamefont {Xiao},
  \citenamefont {Doucet}, \citenamefont {Noh}, \citenamefont {Simmonds},
  \citenamefont {Ranzani}, \citenamefont {Govia},\ and\ \citenamefont
  {Kamal}}]{Xiao2021}%
  \BibitemOpen
  \bibfield  {author} {\bibinfo {author} {\bibfnamefont {Z.}~\bibnamefont
  {Xiao}}, \bibinfo {author} {\bibfnamefont {E.}~\bibnamefont {Doucet}},
  \bibinfo {author} {\bibfnamefont {T.}~\bibnamefont {Noh}}, \bibinfo {author}
  {\bibfnamefont {R.~W.}\ \bibnamefont {Simmonds}}, \bibinfo {author}
  {\bibfnamefont {L.}~\bibnamefont {Ranzani}}, \bibinfo {author} {\bibfnamefont
  {L.~C.~G.}\ \bibnamefont {Govia}}, \ and\ \bibinfo {author} {\bibfnamefont
  {A.}~\bibnamefont {Kamal}},\ }\href@noop {} {\bibfield  {journal} {\bibinfo
  {journal} {Arxiv:}\ }\textbf {\bibinfo {volume} {2103.09260}} (\bibinfo {year}
  {2021})}\BibitemShut {NoStop}%
\bibitem [{\citenamefont {Roch}\ \emph {et~al.}(2014)\citenamefont {Roch},
  \citenamefont {Schwartz}, \citenamefont {Motzoi}, \citenamefont {Macklin},
  \citenamefont {Vijay}, \citenamefont {Eddins}, \citenamefont {Korotkov},
  \citenamefont {Whaley}, \citenamefont {Sarovar},\ and\ \citenamefont
  {Siddiqi}}]{Roch2014}%
  \BibitemOpen
  \bibfield  {author} {\bibinfo {author} {\bibfnamefont {N.}~\bibnamefont
  {Roch}}, \bibinfo {author} {\bibfnamefont {M.~E.}\ \bibnamefont {Schwartz}},
  \bibinfo {author} {\bibfnamefont {F.}~\bibnamefont {Motzoi}}, \bibinfo
  {author} {\bibfnamefont {C.}~\bibnamefont {Macklin}}, \bibinfo {author}
  {\bibfnamefont {R.}~\bibnamefont {Vijay}}, \bibinfo {author} {\bibfnamefont
  {A.~W.}\ \bibnamefont {Eddins}}, \bibinfo {author} {\bibfnamefont {A.~N.}\
  \bibnamefont {Korotkov}}, \bibinfo {author} {\bibfnamefont {K.~B.}\
  \bibnamefont {Whaley}}, \bibinfo {author} {\bibfnamefont {M.}~\bibnamefont
  {Sarovar}}, \ and\ \bibinfo {author} {\bibfnamefont {I.}~\bibnamefont
  {Siddiqi}},\ }\href@noop {} {\bibfield  {journal} {\bibinfo  {journal} {Phys.
  Rev. Lett.}\ }\textbf {\bibinfo {volume} {112}},\ \bibinfo {pages} {170501}
  (\bibinfo {year} {2014})}\BibitemShut {NoStop}%
\bibitem [{\citenamefont {Riste}\ \emph {et~al.}(2013)\citenamefont {Riste},
  \citenamefont {Dukalski}, \citenamefont {Watson}, \citenamefont {de~Lange},
  \citenamefont {Tiggelman}, \citenamefont {Blanter}, \citenamefont {Lehnert},
  \citenamefont {Schouten},\ and\ \citenamefont {DiCarlo}}]{Riste2013}%
  \BibitemOpen
  \bibfield  {author} {\bibinfo {author} {\bibfnamefont {D.}~\bibnamefont
  {Riste}}, \bibinfo {author} {\bibfnamefont {M.}~\bibnamefont {Dukalski}},
  \bibinfo {author} {\bibfnamefont {C.~A.}\ \bibnamefont {Watson}}, \bibinfo
  {author} {\bibfnamefont {G.}~\bibnamefont {de~Lange}}, \bibinfo {author}
  {\bibfnamefont {M.~J.}\ \bibnamefont {Tiggelman}}, \bibinfo {author}
  {\bibfnamefont {Y.~M.}\ \bibnamefont {Blanter}}, \bibinfo {author}
  {\bibfnamefont {K.~W.}\ \bibnamefont {Lehnert}}, \bibinfo {author}
  {\bibfnamefont {R.~N.}\ \bibnamefont {Schouten}}, \ and\ \bibinfo {author}
  {\bibfnamefont {L.}~\bibnamefont {DiCarlo}},\ }\href@noop {} {\bibfield
  {journal} {\bibinfo  {journal} {Nature}\ }\textbf {\bibinfo {volume} {502}},\
  \bibinfo {pages} {350} (\bibinfo {year} {2013})}\BibitemShut {NoStop}%
\bibitem [{\citenamefont {Andersen}\ \emph {et~al.}(2019)\citenamefont
  {Andersen}, \citenamefont {Remm}, \citenamefont {Lazar}, \citenamefont
  {Krinner}, \citenamefont {Heinsoo}, \citenamefont {Besse}, \citenamefont
  {Gabureac}, \citenamefont {Wallraff},\ and\ \citenamefont
  {Eichler}}]{Andersen2019}%
  \BibitemOpen
  \bibfield  {author} {\bibinfo {author} {\bibfnamefont {C.~K.}\ \bibnamefont
  {Andersen}}, \bibinfo {author} {\bibfnamefont {A.}~\bibnamefont {Remm}},
  \bibinfo {author} {\bibfnamefont {S.}~\bibnamefont {Lazar}}, \bibinfo
  {author} {\bibfnamefont {S.}~\bibnamefont {Krinner}}, \bibinfo {author}
  {\bibfnamefont {J.}~\bibnamefont {Heinsoo}}, \bibinfo {author} {\bibfnamefont
  {J.-C.}\ \bibnamefont {Besse}}, \bibinfo {author} {\bibfnamefont
  {M.}~\bibnamefont {Gabureac}}, \bibinfo {author} {\bibfnamefont
  {A.}~\bibnamefont {Wallraff}}, \ and\ \bibinfo {author} {\bibfnamefont
  {C.}~\bibnamefont {Eichler}},\ }\href@noop {} {\bibfield  {journal} {\bibinfo
   {journal} {npj Quantum Information}\ }\textbf {\bibinfo {volume} {5}},\
  \bibinfo {pages} {69} (\bibinfo {year} {2019})}\BibitemShut {NoStop}%
\bibitem [{\citenamefont {Kimchi-Schwartz}\ \emph {et~al.}(2016)\citenamefont
  {Kimchi-Schwartz}, \citenamefont {Martin}, \citenamefont {Flurin},
  \citenamefont {Aron}, \citenamefont {Kulkarni}, \citenamefont {Tureci},\ and\
  \citenamefont {Siddiqi}}]{Schwartz2016}%
  \BibitemOpen
  \bibfield  {author} {\bibinfo {author} {\bibfnamefont {M.~E.}\ \bibnamefont
  {Kimchi-Schwartz}}, \bibinfo {author} {\bibfnamefont {L.}~\bibnamefont
  {Martin}}, \bibinfo {author} {\bibfnamefont {E.}~\bibnamefont {Flurin}},
  \bibinfo {author} {\bibfnamefont {C.}~\bibnamefont {Aron}}, \bibinfo {author}
  {\bibfnamefont {M.}~\bibnamefont {Kulkarni}}, \bibinfo {author}
  {\bibfnamefont {H.~E.}\ \bibnamefont {Tureci}}, \ and\ \bibinfo {author}
  {\bibfnamefont {I.}~\bibnamefont {Siddiqi}},\ }\href@noop {} {\bibfield
  {journal} {\bibinfo  {journal} {Phys. Rev. Lett.}\ }\textbf {\bibinfo
  {volume} {116}},\ \bibinfo {pages} {240503} (\bibinfo {year}
  {2016})}\BibitemShut {NoStop}%
\bibitem [{\citenamefont {Doucet}\ \emph {et~al.}(2020)\citenamefont {Doucet},
  \citenamefont {Reiter}, \citenamefont {Ranzani},\ and\ \citenamefont
  {Kamal}}]{Doucet2020}%
  \BibitemOpen
  \bibfield  {author} {\bibinfo {author} {\bibfnamefont {E.}~\bibnamefont
  {Doucet}}, \bibinfo {author} {\bibfnamefont {F.}~\bibnamefont {Reiter}},
  \bibinfo {author} {\bibfnamefont {L.}~\bibnamefont {Ranzani}}, \ and\
  \bibinfo {author} {\bibfnamefont {A.}~\bibnamefont {Kamal}},\ }\href@noop {}
  {\bibfield  {journal} {\bibinfo  {journal} {Phys. Rev. Research}\ }\textbf
  {\bibinfo {volume} {2}},\ \bibinfo {pages} {023370} (\bibinfo {year}
  {2020})}\BibitemShut {NoStop}%
\bibitem [{\citenamefont {Reagor}\ \emph {et~al.}(2018)\citenamefont {Reagor},
  \citenamefont {Osborn}, \citenamefont {Tezak}, \citenamefont {Staley},
  \citenamefont {Prawiroatmodjo}, \citenamefont {Scheer}, \citenamefont
  {Alidoust}, \citenamefont {Sete}, \citenamefont {Didier}, \citenamefont
  {da~Silva}, \citenamefont {Acala}, \citenamefont {Angeles}, \citenamefont
  {Bestwick}, \citenamefont {Block}, \citenamefont {Bloom}, \citenamefont
  {Bradley}, \citenamefont {Bui}, \citenamefont {Caldwell}, \citenamefont
  {Capelluto}, \citenamefont {Chilcott}, \citenamefont {Cordova}, \citenamefont
  {Crossman}, \citenamefont {Curtis}, \citenamefont {Deshpande}, \citenamefont
  {Bouayadi}, \citenamefont {Girshovich}, \citenamefont {Hong}, \citenamefont
  {Hudson}, \citenamefont {Karalekas}, \citenamefont {Kuang}, \citenamefont
  {Lenihan}, \citenamefont {Manenti}, \citenamefont {Manning}, \citenamefont
  {Marshall}, \citenamefont {Mohan}, \citenamefont {O’Brien}, \citenamefont
  {Otterbach}, \citenamefont {Papageorge}, \citenamefont {Paquette},
  \citenamefont {Pelstring}, \citenamefont {Polloreno}, \citenamefont {Rawat},
  \citenamefont {Ryan}, \citenamefont {Renzas}, \citenamefont {Rubin},
  \citenamefont {Russel}, \citenamefont {Rust}, \citenamefont {Scarabelli},
  \citenamefont {Selvanayagam}, \citenamefont {Sinclair}, \citenamefont
  {Smith}, \citenamefont {Suska}, \citenamefont {To}, \citenamefont
  {Vahidpour}, \citenamefont {Vodrahalli}, \citenamefont {Whyland},
  \citenamefont {Yadav}, \citenamefont {Zeng},\ and\ \citenamefont
  {Rigetti}}]{Reagor2018}%
  \BibitemOpen
  \bibfield  {author} {\bibinfo {author} {\bibfnamefont {M.}~\bibnamefont
  {Reagor}}, \bibinfo {author} {\bibfnamefont {C.~B.}\ \bibnamefont {Osborn}},
  \bibinfo {author} {\bibfnamefont {N.}~\bibnamefont {Tezak}}, \bibinfo
  {author} {\bibfnamefont {A.}~\bibnamefont {Staley}}, \bibinfo {author}
  {\bibfnamefont {G.}~\bibnamefont {Prawiroatmodjo}}, \bibinfo {author}
  {\bibfnamefont {M.}~\bibnamefont {Scheer}}, \bibinfo {author} {\bibfnamefont
  {N.}~\bibnamefont {Alidoust}}, \bibinfo {author} {\bibfnamefont {E.~A.}\
  \bibnamefont {Sete}}, \bibinfo {author} {\bibfnamefont {N.}~\bibnamefont
  {Didier}}, \bibinfo {author} {\bibfnamefont {M.~P.}\ \bibnamefont
  {da~Silva}}, \bibinfo {author} {\bibfnamefont {E.}~\bibnamefont {Acala}},
  \bibinfo {author} {\bibfnamefont {J.}~\bibnamefont {Angeles}}, \bibinfo
  {author} {\bibfnamefont {A.}~\bibnamefont {Bestwick}}, \bibinfo {author}
  {\bibfnamefont {M.}~\bibnamefont {Block}}, \bibinfo {author} {\bibfnamefont
  {B.}~\bibnamefont {Bloom}}, \bibinfo {author} {\bibfnamefont
  {A.}~\bibnamefont {Bradley}}, \bibinfo {author} {\bibfnamefont
  {C.}~\bibnamefont {Bui}}, \bibinfo {author} {\bibfnamefont {S.}~\bibnamefont
  {Caldwell}}, \bibinfo {author} {\bibfnamefont {L.}~\bibnamefont {Capelluto}},
  \bibinfo {author} {\bibfnamefont {R.}~\bibnamefont {Chilcott}}, \bibinfo
  {author} {\bibfnamefont {J.}~\bibnamefont {Cordova}}, \bibinfo {author}
  {\bibfnamefont {G.}~\bibnamefont {Crossman}}, \bibinfo {author}
  {\bibfnamefont {M.}~\bibnamefont {Curtis}}, \bibinfo {author} {\bibfnamefont
  {S.}~\bibnamefont {Deshpande}}, \bibinfo {author} {\bibfnamefont {T.~E.}\
  \bibnamefont {Bouayadi}}, \bibinfo {author} {\bibfnamefont {D.}~\bibnamefont
  {Girshovich}}, \bibinfo {author} {\bibfnamefont {S.}~\bibnamefont {Hong}},
  \bibinfo {author} {\bibfnamefont {A.}~\bibnamefont {Hudson}}, \bibinfo
  {author} {\bibfnamefont {P.}~\bibnamefont {Karalekas}}, \bibinfo {author}
  {\bibfnamefont {K.}~\bibnamefont {Kuang}}, \bibinfo {author} {\bibfnamefont
  {M.}~\bibnamefont {Lenihan}}, \bibinfo {author} {\bibfnamefont
  {R.}~\bibnamefont {Manenti}}, \bibinfo {author} {\bibfnamefont
  {T.}~\bibnamefont {Manning}}, \bibinfo {author} {\bibfnamefont
  {J.}~\bibnamefont {Marshall}}, \bibinfo {author} {\bibfnamefont
  {Y.}~\bibnamefont {Mohan}}, \bibinfo {author} {\bibfnamefont
  {W.}~\bibnamefont {O’Brien}}, \bibinfo {author} {\bibfnamefont
  {J.}~\bibnamefont {Otterbach}}, \bibinfo {author} {\bibfnamefont
  {A.}~\bibnamefont {Papageorge}}, \bibinfo {author} {\bibfnamefont {J.-P.}\
  \bibnamefont {Paquette}}, \bibinfo {author} {\bibfnamefont {M.}~\bibnamefont
  {Pelstring}}, \bibinfo {author} {\bibfnamefont {A.}~\bibnamefont
  {Polloreno}}, \bibinfo {author} {\bibfnamefont {V.}~\bibnamefont {Rawat}},
  \bibinfo {author} {\bibfnamefont {C.~A.}\ \bibnamefont {Ryan}}, \bibinfo
  {author} {\bibfnamefont {R.}~\bibnamefont {Renzas}}, \bibinfo {author}
  {\bibfnamefont {N.}~\bibnamefont {Rubin}}, \bibinfo {author} {\bibfnamefont
  {D.}~\bibnamefont {Russel}}, \bibinfo {author} {\bibfnamefont
  {M.}~\bibnamefont {Rust}}, \bibinfo {author} {\bibfnamefont {D.}~\bibnamefont
  {Scarabelli}}, \bibinfo {author} {\bibfnamefont {M.}~\bibnamefont
  {Selvanayagam}}, \bibinfo {author} {\bibfnamefont {R.}~\bibnamefont
  {Sinclair}}, \bibinfo {author} {\bibfnamefont {R.}~\bibnamefont {Smith}},
  \bibinfo {author} {\bibfnamefont {M.}~\bibnamefont {Suska}}, \bibinfo
  {author} {\bibfnamefont {T.-W.}\ \bibnamefont {To}}, \bibinfo {author}
  {\bibfnamefont {M.}~\bibnamefont {Vahidpour}}, \bibinfo {author}
  {\bibfnamefont {N.}~\bibnamefont {Vodrahalli}}, \bibinfo {author}
  {\bibfnamefont {T.}~\bibnamefont {Whyland}}, \bibinfo {author} {\bibfnamefont
  {K.}~\bibnamefont {Yadav}}, \bibinfo {author} {\bibfnamefont
  {W.}~\bibnamefont {Zeng}}, \ and\ \bibinfo {author} {\bibfnamefont {C.~T.}\
  \bibnamefont {Rigetti}},\ }\href {\doibase
  https://doi.org/10.1126/sciadv.aao3603} {\bibfield  {journal} {\bibinfo
  {journal} {Science Advances}\ }\textbf {\bibinfo {volume} {4}},\ \bibinfo
  {pages} {2} (\bibinfo {year} {2018})}\BibitemShut {NoStop}%
\bibitem [{\citenamefont {Noguchi}\ \emph {et~al.}(2018)\citenamefont
  {Noguchi}, \citenamefont {Osada}, \citenamefont {Masuda}, \citenamefont
  {Kono}, \citenamefont {Heya}, \citenamefont {Wolski}, \citenamefont
  {Takahashi}, \citenamefont {Sugiyama}, \citenamefont {Lachance-Quirion},\
  and\ \citenamefont {Nakamura}}]{Noguchi2018}%
  \BibitemOpen
  \bibfield  {author} {\bibinfo {author} {\bibfnamefont {A.}~\bibnamefont
  {Noguchi}}, \bibinfo {author} {\bibfnamefont {A.}~\bibnamefont {Osada}},
  \bibinfo {author} {\bibfnamefont {S.}~\bibnamefont {Masuda}}, \bibinfo
  {author} {\bibfnamefont {S.}~\bibnamefont {Kono}}, \bibinfo {author}
  {\bibfnamefont {K.}~\bibnamefont {Heya}}, \bibinfo {author} {\bibfnamefont
  {S.~P.}\ \bibnamefont {Wolski}}, \bibinfo {author} {\bibfnamefont
  {H.}~\bibnamefont {Takahashi}}, \bibinfo {author} {\bibfnamefont
  {T.}~\bibnamefont {Sugiyama}}, \bibinfo {author} {\bibfnamefont
  {D.}~\bibnamefont {Lachance-Quirion}}, \ and\ \bibinfo {author}
  {\bibfnamefont {Y.}~\bibnamefont {Nakamura}},\ }\href@noop {} {\bibfield
  {journal} {\bibinfo  {journal} {Phys. Rev. A}\ }\textbf {\bibinfo {volume}
  {102}},\ \bibinfo {pages} {062408} (\bibinfo {year} {2018})}\BibitemShut
  {NoStop}%
\bibitem [{\citenamefont {Blais}\ \emph {et~al.}(2004)\citenamefont {Blais},
  \citenamefont {Huang}, \citenamefont {Wallraff}, \citenamefont {Girvin},\
  and\ \citenamefont {Schoelkopf}}]{schoelkopf2004}%
  \BibitemOpen
  \bibfield  {author} {\bibinfo {author} {\bibfnamefont {A.}~\bibnamefont
  {Blais}}, \bibinfo {author} {\bibfnamefont {R.-S.}\ \bibnamefont {Huang}},
  \bibinfo {author} {\bibfnamefont {A.}~\bibnamefont {Wallraff}}, \bibinfo
  {author} {\bibfnamefont {S.~M.}\ \bibnamefont {Girvin}}, \ and\ \bibinfo
  {author} {\bibfnamefont {R.~J.}\ \bibnamefont {Schoelkopf}},\ }\href@noop {}
  {\bibfield  {journal} {\bibinfo  {journal} {Phys. Rev. A}\ }\textbf {\bibinfo
  {volume} {69}},\ \bibinfo {pages} {062320} (\bibinfo {year}
  {2004})}\BibitemShut {NoStop}%
\bibitem [{\citenamefont {Koch}\ \emph {et~al.}(2007)\citenamefont {Koch},
  \citenamefont {Yu}, \citenamefont {Gambetta}, \citenamefont {Houck},
  \citenamefont {Schuster}, \citenamefont {Majer}, \citenamefont {Blais},
  \citenamefont {Devoret}, \citenamefont {Girvin},\ and\ \citenamefont
  {Schoelkopf}}]{Koch2007}%
  \BibitemOpen
  \bibfield  {author} {\bibinfo {author} {\bibfnamefont {J.}~\bibnamefont
  {Koch}}, \bibinfo {author} {\bibfnamefont {T.~M.}\ \bibnamefont {Yu}},
  \bibinfo {author} {\bibfnamefont {J.}~\bibnamefont {Gambetta}}, \bibinfo
  {author} {\bibfnamefont {A.~A.}\ \bibnamefont {Houck}}, \bibinfo {author}
  {\bibfnamefont {D.~I.}\ \bibnamefont {Schuster}}, \bibinfo {author}
  {\bibfnamefont {J.}~\bibnamefont {Majer}}, \bibinfo {author} {\bibfnamefont
  {A.}~\bibnamefont {Blais}}, \bibinfo {author} {\bibfnamefont {M.~H.}\
  \bibnamefont {Devoret}}, \bibinfo {author} {\bibfnamefont {S.~M.}\
  \bibnamefont {Girvin}}, \ and\ \bibinfo {author} {\bibfnamefont {R.~J.}\
  \bibnamefont {Schoelkopf}},\ }\href@noop {} {\bibfield  {journal} {\bibinfo
  {journal} {Phys. Rev. A}\ }\textbf {\bibinfo {volume} {76}},\ \bibinfo
  {pages} {042319} (\bibinfo {year} {2007})}\BibitemShut {NoStop}%
\bibitem [{\citenamefont {Schuster}\ \emph {et~al.}(2005)\citenamefont
  {Schuster}, \citenamefont {Wallraff}, \citenamefont {Blais}, \citenamefont
  {Frunzio}, \citenamefont {Huang}, \citenamefont {Majer}, \citenamefont
  {Girvin},\ and\ \citenamefont {Schoelkopf}}]{StarkShift2005}%
  \BibitemOpen
  \bibfield  {author} {\bibinfo {author} {\bibfnamefont {D.~I.}\ \bibnamefont
  {Schuster}}, \bibinfo {author} {\bibfnamefont {A.}~\bibnamefont {Wallraff}},
  \bibinfo {author} {\bibfnamefont {A.}~\bibnamefont {Blais}}, \bibinfo
  {author} {\bibfnamefont {L.}~\bibnamefont {Frunzio}}, \bibinfo {author}
  {\bibfnamefont {R.-S.}\ \bibnamefont {Huang}}, \bibinfo {author}
  {\bibfnamefont {J.}~\bibnamefont {Majer}}, \bibinfo {author} {\bibfnamefont
  {S.}~\bibnamefont {Girvin}}, \ and\ \bibinfo {author} {\bibfnamefont {R.~J.}\
  \bibnamefont {Schoelkopf}},\ }\href@noop {} {\bibfield  {journal} {\bibinfo
  {journal} {Phys. Rev. Lett.}\ }\textbf {\bibinfo {volume} {94}},\ \bibinfo
  {pages} {123602} (\bibinfo {year} {2005})}\BibitemShut {NoStop}%
\bibitem [{\citenamefont {Houck}\ \emph {et~al.}(2008)\citenamefont {Houck},
  \citenamefont {Schreier}, \citenamefont {Johnson}, \citenamefont {Chow},
  \citenamefont {Koch}, \citenamefont {Gambetta}, \citenamefont {Schuster},
  \citenamefont {Frunzio}, \citenamefont {Devoret}, \citenamefont {Girvin},\
  and\ \citenamefont {Schoelkopf}}]{Houck2008}%
  \BibitemOpen
  \bibfield  {author} {\bibinfo {author} {\bibfnamefont {A.~A.}\ \bibnamefont
  {Houck}}, \bibinfo {author} {\bibfnamefont {J.~A.}\ \bibnamefont {Schreier}},
  \bibinfo {author} {\bibfnamefont {B.~R.}\ \bibnamefont {Johnson}}, \bibinfo
  {author} {\bibfnamefont {J.~M.}\ \bibnamefont {Chow}}, \bibinfo {author}
  {\bibfnamefont {J.}~\bibnamefont {Koch}}, \bibinfo {author} {\bibfnamefont
  {J.~M.}\ \bibnamefont {Gambetta}}, \bibinfo {author} {\bibfnamefont {D.~I.}\
  \bibnamefont {Schuster}}, \bibinfo {author} {\bibfnamefont {L.}~\bibnamefont
  {Frunzio}}, \bibinfo {author} {\bibfnamefont {M.~H.}\ \bibnamefont
  {Devoret}}, \bibinfo {author} {\bibfnamefont {S.~M.}\ \bibnamefont {Girvin}},
  \ and\ \bibinfo {author} {\bibfnamefont {R.~J.}\ \bibnamefont {Schoelkopf}},\
  }\href@noop {} {\bibfield  {journal} {\bibinfo  {journal} {Phys. Rev. Lett.}\
  }\textbf {\bibinfo {volume} {101}},\ \bibinfo {pages} {080502} (\bibinfo
  {year} {2008})}\BibitemShut {NoStop}%
\bibitem [{\citenamefont {Chen}\ \emph {et~al.}(2014)\citenamefont {Chen},
  \citenamefont {Neill}, \citenamefont {Roushan}, \citenamefont {Leung},
  \citenamefont {Fang}, \citenamefont {Barends}, \citenamefont {J.~Kelly},
  \citenamefont {Z.~Chen}, \citenamefont {Dunsworth}, \citenamefont {Jeffrey},
  \citenamefont {Megrant}, \citenamefont {Mutus}, \citenamefont {O’Malley},
  \citenamefont {Quintana}, \citenamefont {Sank}, \citenamefont {Vainsencher},
  \citenamefont {Wenner}, \citenamefont {White}, \citenamefont {Geller},
  \citenamefont {Cleland},\ and\ \citenamefont {Martinis}}]{Chen2014}%
  \BibitemOpen
  \bibfield  {author} {\bibinfo {author} {\bibfnamefont {Y.}~\bibnamefont
  {Chen}}, \bibinfo {author} {\bibfnamefont {C.}~\bibnamefont {Neill}},
  \bibinfo {author} {\bibfnamefont {P.}~\bibnamefont {Roushan}}, \bibinfo
  {author} {\bibfnamefont {N.}~\bibnamefont {Leung}}, \bibinfo {author}
  {\bibfnamefont {M.}~\bibnamefont {Fang}}, \bibinfo {author} {\bibfnamefont
  {R.}~\bibnamefont {Barends}}, \bibinfo {author} {\bibfnamefont {B.~C.}\
  \bibnamefont {J.~Kelly}}, \bibinfo {author} {\bibfnamefont {B.~C.}\
  \bibnamefont {Z.~Chen}}, \bibinfo {author} {\bibfnamefont {A.}~\bibnamefont
  {Dunsworth}}, \bibinfo {author} {\bibfnamefont {E.}~\bibnamefont {Jeffrey}},
  \bibinfo {author} {\bibfnamefont {A.}~\bibnamefont {Megrant}}, \bibinfo
  {author} {\bibfnamefont {J.~Y.}\ \bibnamefont {Mutus}}, \bibinfo {author}
  {\bibfnamefont {P.~J.~J.}\ \bibnamefont {O’Malley}}, \bibinfo {author}
  {\bibfnamefont {C.~M.}\ \bibnamefont {Quintana}}, \bibinfo {author}
  {\bibfnamefont {D.}~\bibnamefont {Sank}}, \bibinfo {author} {\bibfnamefont
  {A.}~\bibnamefont {Vainsencher}}, \bibinfo {author} {\bibfnamefont
  {J.}~\bibnamefont {Wenner}}, \bibinfo {author} {\bibfnamefont {T.~C.}\
  \bibnamefont {White}}, \bibinfo {author} {\bibfnamefont {M.~R.}\ \bibnamefont
  {Geller}}, \bibinfo {author} {\bibfnamefont {A.~N.}\ \bibnamefont {Cleland}},
  \ and\ \bibinfo {author} {\bibfnamefont {J.~M.}\ \bibnamefont {Martinis}},\
  }\href@noop {} {\bibfield  {journal} {\bibinfo  {journal} {Phys. Rev. Lett.}\
  }\textbf {\bibinfo {volume} {113}},\ \bibinfo {pages} {220502} (\bibinfo
  {year} {2014})}\BibitemShut {NoStop}%
\bibitem [{\citenamefont {Roushan}\ \emph {et~al.}(2017)\citenamefont
  {Roushan}, \citenamefont {Neill}, \citenamefont {Megrant}, \citenamefont
  {Chen}, \citenamefont {Babbush}, \citenamefont {Barends}, \citenamefont
  {Campbell}, \citenamefont {Chen}, \citenamefont {Chiaro}, \citenamefont
  {Dunsworth}, \citenamefont {Fowler}, \citenamefont {Jeffrey}, \citenamefont
  {Kelly}, \citenamefont {Lucero}, \citenamefont {Mutus}, \citenamefont
  {O’Malley}, \citenamefont {Neeley}, \citenamefont {Quintana}, \citenamefont
  {Sank}, \citenamefont {Vainsencher}, \citenamefont {J.Wenner}, \citenamefont
  {White}, \citenamefont {Kapit}, \citenamefont {Neven},\ and\ \citenamefont
  {Martinis}}]{Roushan2017}%
  \BibitemOpen
  \bibfield  {author} {\bibinfo {author} {\bibfnamefont {P.}~\bibnamefont
  {Roushan}}, \bibinfo {author} {\bibfnamefont {C.}~\bibnamefont {Neill}},
  \bibinfo {author} {\bibfnamefont {A.}~\bibnamefont {Megrant}}, \bibinfo
  {author} {\bibfnamefont {Y.}~\bibnamefont {Chen}}, \bibinfo {author}
  {\bibfnamefont {R.}~\bibnamefont {Babbush}}, \bibinfo {author} {\bibfnamefont
  {R.}~\bibnamefont {Barends}}, \bibinfo {author} {\bibfnamefont
  {B.}~\bibnamefont {Campbell}}, \bibinfo {author} {\bibfnamefont
  {Z.}~\bibnamefont {Chen}}, \bibinfo {author} {\bibfnamefont {B.}~\bibnamefont
  {Chiaro}}, \bibinfo {author} {\bibfnamefont {A.}~\bibnamefont {Dunsworth}},
  \bibinfo {author} {\bibfnamefont {A.}~\bibnamefont {Fowler}}, \bibinfo
  {author} {\bibfnamefont {E.}~\bibnamefont {Jeffrey}}, \bibinfo {author}
  {\bibfnamefont {J.}~\bibnamefont {Kelly}}, \bibinfo {author} {\bibfnamefont
  {E.}~\bibnamefont {Lucero}}, \bibinfo {author} {\bibfnamefont
  {J.}~\bibnamefont {Mutus}}, \bibinfo {author} {\bibfnamefont {P.~J.~J.}\
  \bibnamefont {O’Malley}}, \bibinfo {author} {\bibfnamefont
  {M.}~\bibnamefont {Neeley}}, \bibinfo {author} {\bibfnamefont
  {C.}~\bibnamefont {Quintana}}, \bibinfo {author} {\bibfnamefont
  {D.}~\bibnamefont {Sank}}, \bibinfo {author} {\bibfnamefont {A.}~\bibnamefont
  {Vainsencher}}, \bibinfo {author} {\bibnamefont {J.Wenner}}, \bibinfo
  {author} {\bibfnamefont {T.}~\bibnamefont {White}}, \bibinfo {author}
  {\bibfnamefont {E.}~\bibnamefont {Kapit}}, \bibinfo {author} {\bibfnamefont
  {H.}~\bibnamefont {Neven}}, \ and\ \bibinfo {author} {\bibfnamefont
  {J.}~\bibnamefont {Martinis}},\ }\href@noop {} {\bibfield  {journal}
  {\bibinfo  {journal} {Nature Physics}\ }\textbf {\bibinfo {volume} {13}},\
  \bibinfo {pages} {146} (\bibinfo {year} {2017})}\BibitemShut {NoStop}%
\bibitem [{\citenamefont {Nguyen}\ \emph {et~al.}(2012)\citenamefont {Nguyen},
  \citenamefont {Zakka-Bajjani}, \citenamefont {Simmonds},\ and\ \citenamefont
  {Aumentado}}]{Nguyen2012}%
  \BibitemOpen
  \bibfield  {author} {\bibinfo {author} {\bibfnamefont {F.}~\bibnamefont
  {Nguyen}}, \bibinfo {author} {\bibfnamefont {E.}~\bibnamefont
  {Zakka-Bajjani}}, \bibinfo {author} {\bibfnamefont {R.~W.}\ \bibnamefont
  {Simmonds}}, \ and\ \bibinfo {author} {\bibfnamefont {J.}~\bibnamefont
  {Aumentado}},\ }\href@noop {} {\bibfield  {journal} {\bibinfo  {journal}
  {Phys. Rev. Lett.}\ }\textbf {\bibinfo {volume} {108}},\ \bibinfo {pages}
  {163602} (\bibinfo {year} {2012})}\BibitemShut {NoStop}%
\bibitem [{\citenamefont {Rosenblum}\ \emph {et~al.}(2018)\citenamefont
  {Rosenblum}, \citenamefont {Gao}, \citenamefont {Reinhold}, \citenamefont
  {Wang}, \citenamefont {Axline}, \citenamefont {Frunzio}, \citenamefont
  {Girvin}, \citenamefont {Jiang}, \citenamefont {Mirrahimi}, \citenamefont
  {Devoret},\ and\ \citenamefont {Schoelkopf}}]{Rosenblum2018a}%
  \BibitemOpen
  \bibfield  {author} {\bibinfo {author} {\bibfnamefont {S.}~\bibnamefont
  {Rosenblum}}, \bibinfo {author} {\bibfnamefont {Y.}~\bibnamefont {Gao}},
  \bibinfo {author} {\bibfnamefont {P.}~\bibnamefont {Reinhold}}, \bibinfo
  {author} {\bibfnamefont {C.}~\bibnamefont {Wang}}, \bibinfo {author}
  {\bibfnamefont {C.}~\bibnamefont {Axline}}, \bibinfo {author} {\bibfnamefont
  {L.}~\bibnamefont {Frunzio}}, \bibinfo {author} {\bibfnamefont
  {S.}~\bibnamefont {Girvin}}, \bibinfo {author} {\bibfnamefont
  {L.}~\bibnamefont {Jiang}}, \bibinfo {author} {\bibfnamefont
  {M.}~\bibnamefont {Mirrahimi}}, \bibinfo {author} {\bibfnamefont
  {M.}~\bibnamefont {Devoret}}, \ and\ \bibinfo {author} {\bibfnamefont
  {R.}~\bibnamefont {Schoelkopf}},\ }\href@noop {} {\bibfield  {journal}
  {\bibinfo  {journal} {Nature Communications}\ }\textbf {\bibinfo {volume}
  {9}},\ \bibinfo {pages} {652} (\bibinfo {year} {2018})}\BibitemShut {NoStop}%
\bibitem [{\citenamefont {Rosenblum}\ \emph {et~al.}(2012)\citenamefont
  {Rosenblum}, \citenamefont {Reinhold}, \citenamefont {Mirrahimi},
  \citenamefont {Jiang}, \citenamefont {Frunzio},\ and\ \citenamefont
  {Schoelkopf}}]{Rosenblum2018b}%
  \BibitemOpen
  \bibfield  {author} {\bibinfo {author} {\bibfnamefont {S.}~\bibnamefont
  {Rosenblum}}, \bibinfo {author} {\bibfnamefont {P.}~\bibnamefont {Reinhold}},
  \bibinfo {author} {\bibfnamefont {M.}~\bibnamefont {Mirrahimi}}, \bibinfo
  {author} {\bibfnamefont {L.}~\bibnamefont {Jiang}}, \bibinfo {author}
  {\bibfnamefont {L.}~\bibnamefont {Frunzio}}, \ and\ \bibinfo {author}
  {\bibfnamefont {R.}~\bibnamefont {Schoelkopf}},\ }\href@noop {} {\bibfield
  {journal} {\bibinfo  {journal} {Science}\ }\textbf {\bibinfo {volume}
  {361}},\ \bibinfo {pages} {266} (\bibinfo {year} {2012})}\BibitemShut
  {NoStop}%
\bibitem [{\citenamefont {{\it et al.}}(2021{\natexlab{a}})}]{Brown2021}%
  \BibitemOpen
  \bibfield  {author} {\bibinfo {author} {\bibfnamefont {T.~B.}\ \bibnamefont
  {{\it et al.}}},\ }\href@noop {} {\bibfield  {journal} {\bibinfo  {journal}
  {Arxiv:}\ }\textbf {\bibinfo {volume} {{\it in preparation}}} (\bibinfo
  {year} {2021}{\natexlab{a}})}\BibitemShut {NoStop}%
\bibitem [{\citenamefont {{\it et al.}}(2021{\natexlab{b}})}]{Jin2021a}%
  \BibitemOpen
  \bibfield  {author} {\bibinfo {author} {\bibfnamefont {X.~Y.~J.}\
  \bibnamefont {{\it et al.}}},\ }\href@noop {} {\bibfield  {journal} {\bibinfo
   {journal} {Arxiv:}\ }\textbf {\bibinfo {volume} {{\it in preparation}}}
  (\bibinfo {year} {2021}{\natexlab{b}})}\BibitemShut {NoStop}%
\bibitem [{\citenamefont {{\it et al.}}(2021{\natexlab{c}})}]{Jin2021b}%
  \BibitemOpen
  \bibfield  {author} {\bibinfo {author} {\bibfnamefont {X.~Y.~J.}\
  \bibnamefont {{\it et al.}}},\ }\href@noop {} {\bibfield  {journal} {\bibinfo
   {journal} {Arxiv:}\ }\textbf {\bibinfo {volume} {{\it in preparation}}}
  (\bibinfo {year} {2021}{\natexlab{c}})}\BibitemShut {NoStop}%
\bibitem [{\citenamefont {Boissonneault}\ \emph {et~al.}(2009)\citenamefont
  {Boissonneault}, \citenamefont {Gambetta},\ and\ \citenamefont
  {Blais}}]{blais2009}%
  \BibitemOpen
  \bibfield  {author} {\bibinfo {author} {\bibfnamefont {M.}~\bibnamefont
  {Boissonneault}}, \bibinfo {author} {\bibfnamefont {J.~M.}\ \bibnamefont
  {Gambetta}}, \ and\ \bibinfo {author} {\bibfnamefont {A.}~\bibnamefont
  {Blais}},\ }\href@noop {} {\bibfield  {journal} {\bibinfo  {journal} {Phys.
  Rev. A}\ }\textbf {\bibinfo {volume} {79}},\ \bibinfo {pages} {013819}
  (\bibinfo {year} {2009})}\BibitemShut {NoStop}%
\bibitem [{\citenamefont {Gambetta}\ \emph {et~al.}(2006)\citenamefont
  {Gambetta}, \citenamefont {Blais}, \citenamefont {Schuster}, \citenamefont
  {Wallraff}, \citenamefont {L.~Frunzio}, \citenamefont {Devoret},
  \citenamefont {Girvin},\ and\ \citenamefont {Schoelkopf}}]{Gambetta2006}%
  \BibitemOpen
  \bibfield  {author} {\bibinfo {author} {\bibfnamefont {J.}~\bibnamefont
  {Gambetta}}, \bibinfo {author} {\bibfnamefont {A.}~\bibnamefont {Blais}},
  \bibinfo {author} {\bibfnamefont {D.~I.}\ \bibnamefont {Schuster}}, \bibinfo
  {author} {\bibfnamefont {A.}~\bibnamefont {Wallraff}}, \bibinfo {author}
  {\bibfnamefont {J.~M.}\ \bibnamefont {L.~Frunzio}}, \bibinfo {author}
  {\bibfnamefont {M.~H.}\ \bibnamefont {Devoret}}, \bibinfo {author}
  {\bibfnamefont {S.~M.}\ \bibnamefont {Girvin}}, \ and\ \bibinfo {author}
  {\bibfnamefont {R.~J.}\ \bibnamefont {Schoelkopf}},\ }\href@noop {}
  {\bibfield  {journal} {\bibinfo  {journal} {Phys. Rev. A}\ }\textbf {\bibinfo
  {volume} {74}},\ \bibinfo {pages} {042318} (\bibinfo {year}
  {2006})}\BibitemShut {NoStop}%
\bibitem [{\citenamefont {Reed}\ \emph {et~al.}(2010)\citenamefont {Reed},
  \citenamefont {Johnson}, \citenamefont {Houck}, \citenamefont {DiCarlo},
  \citenamefont {Chow}, \citenamefont {Schuster}, \citenamefont {Frunzio},\
  and\ \citenamefont {Schoelkopf}}]{Reed2010}%
  \BibitemOpen
  \bibfield  {author} {\bibinfo {author} {\bibfnamefont {M.~D.}\ \bibnamefont
  {Reed}}, \bibinfo {author} {\bibfnamefont {B.~R.}\ \bibnamefont {Johnson}},
  \bibinfo {author} {\bibfnamefont {A.~A.}\ \bibnamefont {Houck}}, \bibinfo
  {author} {\bibfnamefont {L.}~\bibnamefont {DiCarlo}}, \bibinfo {author}
  {\bibfnamefont {J.~M.}\ \bibnamefont {Chow}}, \bibinfo {author}
  {\bibfnamefont {D.~I.}\ \bibnamefont {Schuster}}, \bibinfo {author}
  {\bibfnamefont {L.}~\bibnamefont {Frunzio}}, \ and\ \bibinfo {author}
  {\bibfnamefont {R.~J.}\ \bibnamefont {Schoelkopf}},\ }\href@noop {}
  {\bibfield  {journal} {\bibinfo  {journal} {Phys. Rev. Lett.}\ }\textbf
  {\bibinfo {volume} {96}},\ \bibinfo {pages} {203110} (\bibinfo {year}
  {2010})}\BibitemShut {NoStop}%
\end{thebibliography}
\end{document}


\newcommand{\ms}[1]{\mbox{\scriptsize #1}}
\newcommand{\msb}[1]{\mbox{\scriptsize $\mathbf{#1}$}}
\newcommand{\msi}[1]{\mbox{\scriptsize\textit{#1}}}
\newcommand{\nn}{\nonumber} 
\newcommand{\dg}{^\dagger}
\newcommand{\smallfrac}[2]{\mbox{$\frac{#1}{#2}$}}
\newcommand{\pfpx}[2]{\frac{\partial #1}{\partial #2}}
\newcommand{\dfdx}[2]{\frac{d #1}{d #2}}
\newcommand{\half}{\smallfrac{1}{2}}
\newcommand{\s}{{\mathcal S}}
\newcommand{\red}{\color{red}}
\newcommand{\bluetext}{\color{blue}}
\newcommand{\rws}{\color{blue}}
\newcommand{\kurt}{\color{green}} 
\newtheorem{theo}{Theorem} \newtheorem{lemma}{Lemma}

\title{Supplemental Material: Strong parametric dispersive shifts in a statically decoupled multi-qubit cavity QED system}

\author{T. Noh}
\email{taewan.noh@nist.gov}
\affiliation{Associate of the National Institute of Standards and Technology, Boulder, Colorado 80305, USA}
\affiliation{Department of Physics and Applied Physics, University of Massachusetts, Lowell, MA 01854, USA}
\author{Z. Xiao}
\affiliation{Department of Physics and Applied Physics, University of Massachusetts, Lowell, MA 01854, USA}
\author{K. Cicak}
\affiliation{National Institute of Standards and Technology, 325 Broadway St MS686.05, Boulder, Colorado 80305, USA}
\author{X. Y. Jin}
\affiliation{Associate of the National Institute of Standards and Technology, Boulder, Colorado 80305, USA}
\affiliation{Department of Physics, University of Colorado, Boulder, Colorado 80309, USA}
\author{E. Doucet}
\affiliation{Department of Physics and Applied Physics, University of Massachusetts, Lowell, MA 01854, USA}
\author{J. Teufel}
\affiliation{National Institute of Standards and Technology, 325 Broadway St MS686.05, Boulder, Colorado 80305, USA}
\author{J. Aumentado}
\affiliation{National Institute of Standards and Technology, 325 Broadway St MS686.05, Boulder, Colorado 80305, USA}
\author{L. C. G. Govia}
\affiliation{Quantum Engineering and Computing, Raytheon BBN Technologies, Cambridge, Massachusetts 02138, USA}
\author{L. Ranzani}
\affiliation{Quantum Engineering and Computing, Raytheon BBN Technologies, Cambridge, Massachusetts 02138, USA}
\author{A. Kamal}
\affiliation{Department of Physics and Applied Physics, University of Massachusetts, Lowell, MA 01854, USA}
\author{R. W. Simmonds}
\email{raymond.simmonds@nist.gov}
\affiliation{National Institute of Standards and Technology, 325 Broadway St MS686.05, Boulder, Colorado 80305, USA}
\date{\today}%
\maketitle

\section{Calibration of the parametric pump drive}

\begin{figure}[t]
\centering
\includegraphics[width=10cm]{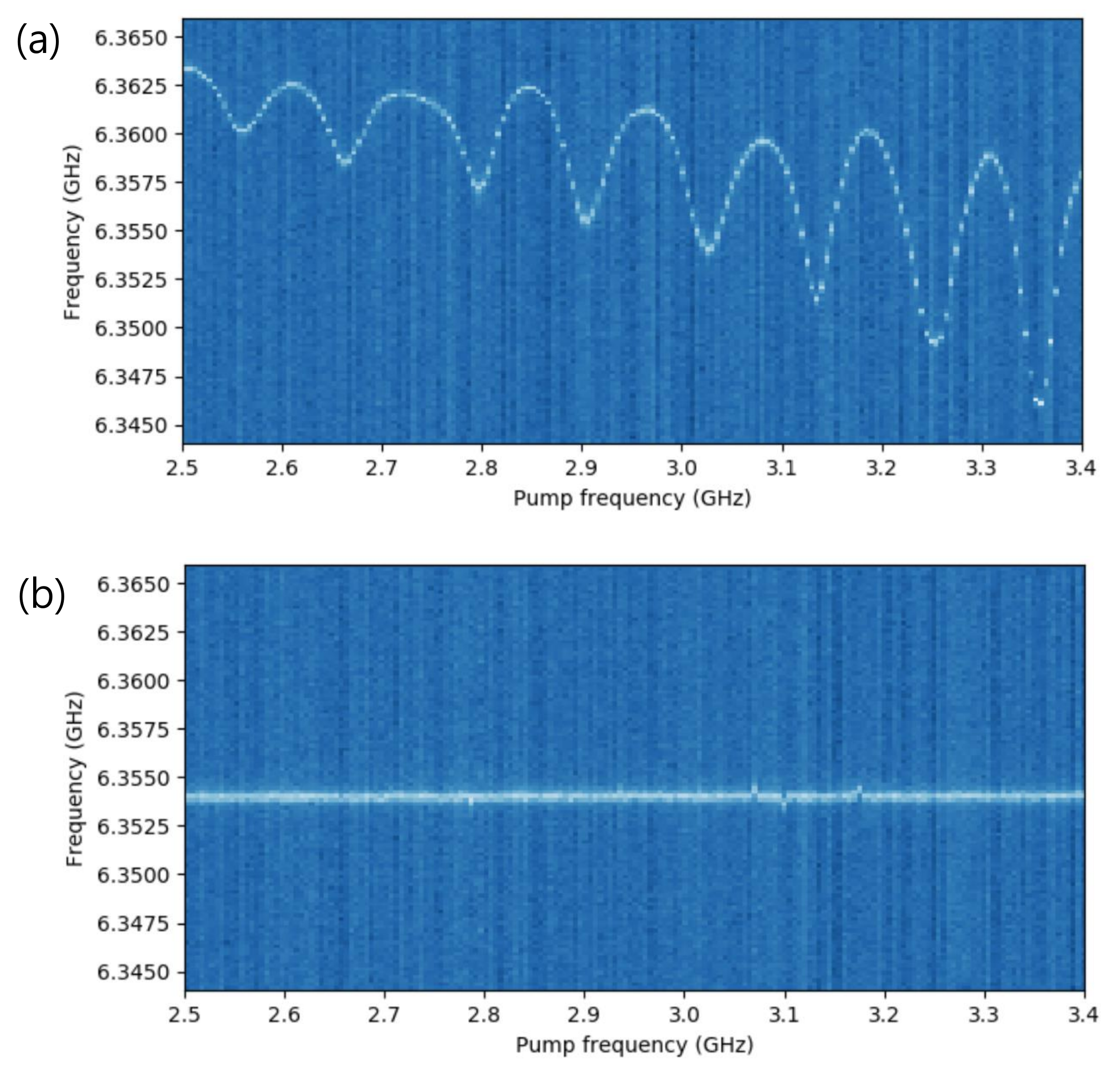}%
\caption{Resonant frequency of the R transmon as a function of the frequency of parametric pump: (a) before calibration  and (b) after calibration of the amplitude of parametric pump drive. The shift of resonant frequency of the R transmon at each pump frequency is proportional to the square of pump amplitude $\delta \phi_p$.}
\end{figure}

Fig. 2(a) in the main exhibits a typical cavity spectrum while sweeping the frequency $\omega_p$ of parametric pump. During the measurement, it is crucial to maintain the amplitude of the pump as a constant. In this section, we discuss a systematic procedure to calibrate the amplitude of parametric pump.

In the presence of parametric pump, the resonant frequency of the transmon $\omega_k (\phi)$ $(k=L, R)$ at a  given flux bias $\phi = \Phi/\Phi_0$ is shifted to an average value $\overline{\omega_k (\phi)}$. If the flux is modulated with an amplitude $\delta \phi_p$ with respect to $\phi_s$, $\overline{\omega_k (\phi)}$ can be expressed as
\begin{eqnarray}
\overline{\omega_k (\phi)} &=& \left(\int^{\phi_s + \frac{1}{2} \delta \phi_p}_{\phi_s - \frac{1}{2} \delta \phi_p} \omega_k (\phi) d\phi \right)/\delta \phi_p \\ \nonumber
&=& \omega_k (\phi_s)+ \frac{1}{24} \omega_k ''(\phi_s) \cdot(\delta \phi_p)^2 + \cdots.
\end{eqnarray}
If we ignore the contributions of higher order terms of $\delta \phi_p$, the result indicates that the shift $\delta \omega_k = \overline{\omega_k (\phi)} - \omega_k (\phi_s)$ is proportional to the second derivative of $\omega_k (\phi)$ at $\phi_s$ and the square of the amplitude of the pump. In other words, the relative amplitude of the pump in the range of sweeping frequency can be extracted by measuring the shift of the resonant frequency of either transmon from its value in the absence of the pump, at a fixed flux bias.

To be more specific, we measured the resonant frequency of the R transmon at $\phi_s = 0$ while sweeping the pump frequency between 2.5 GHz and 3.4 GHz at a fixed amplitude. The flux bias to perform the measurement has been chosen on purpose where $\omega_R'(\phi_s)=0$ in order to suppress any unwanted parametric coupling between the transmon and cavity, which is known to be proportional to $\sqrt{(\frac{d\omega_R}{d\phi})(\frac{d\omega_C}{d\phi}})$. In this way, we can measure the shift in the resonant frequency that is purely dependent on the power of the pump, i.e., $(\delta \phi)^2$. Fig. S1(a) shows the resonant frequency of R transmon, which is highly dependent on the frequency of parametric pump. Hence, this result indicates that the actual amplitude of the pump varies significantly in the range of the swept frequency. Therefore, at each frequency, a numerical factor 
\begin{equation}
\lambda = \lambda_0/\sqrt{|\overline{\omega_R (\phi_s)} - \omega_R (\phi_s)|}    
\end{equation}
should be multiplied to maintain a constant amplitude, where $\lambda_0$ is determined by the choice of the frequency of parametric pump at which we would like to normalize $\lambda$. Fig. S1(b) shows the result of the same measurement as (a), but with the numerical factor $\lambda$ for the amplitude taken into account. Here, we have chosen 3.03 GHz of pump frequency to normalize $\lambda$. The resonant frequency of the R transmon is now nearly constant in the range of the swept frequency, which clearly demonstrates the calibration of the amplitude of parametric pump.

\begin{figure}[t]
\centering
\includegraphics[width=13cm]{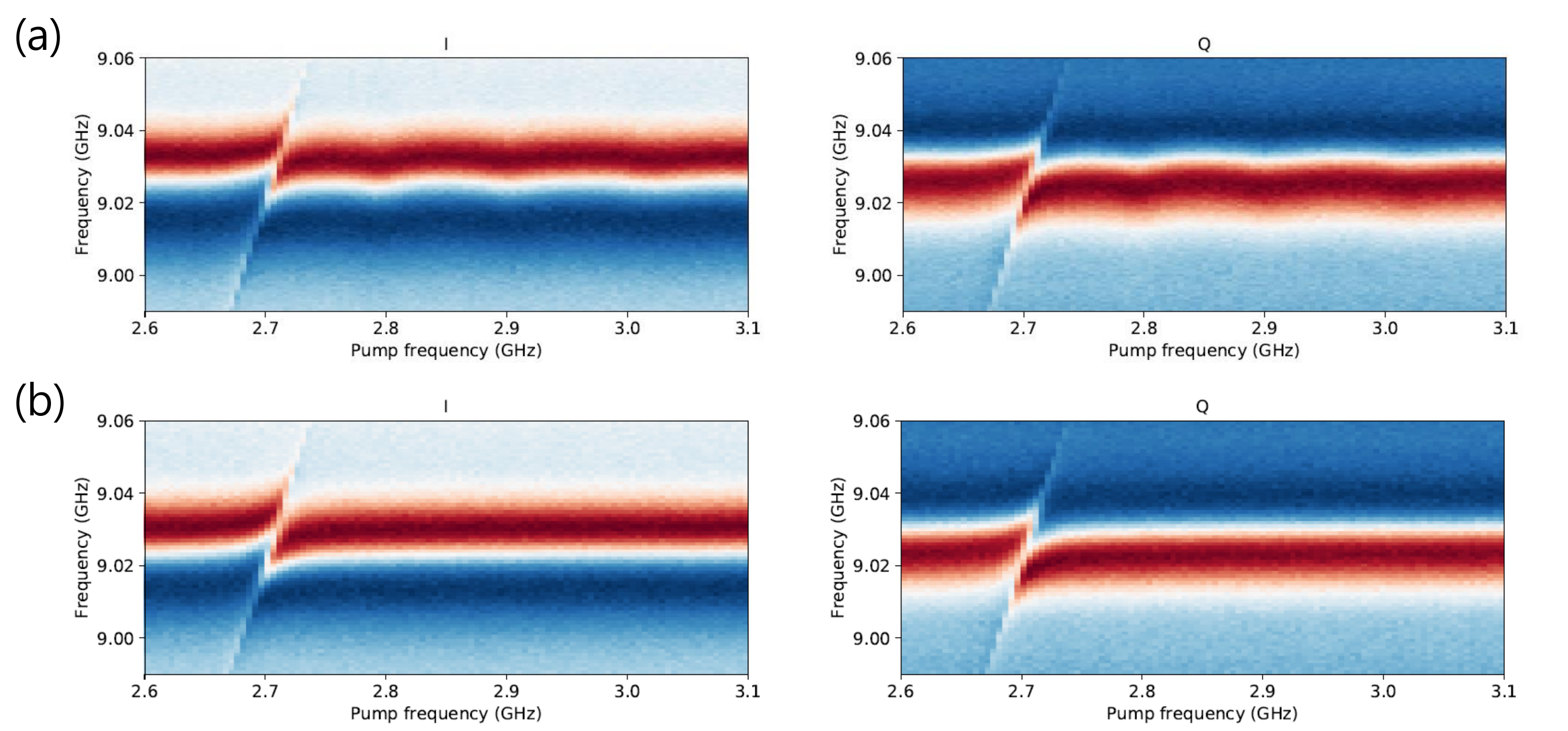}%
\caption{In-phase (I) and out-of-phase (Q) quadratures of the cavity spectrum measured near the resonant frequency of the cavity as a function of the frequency of parametric pump: (a) I and Q before the calibration, (b) I and Q after the calibration of the amplitude of parametric pump.}
\end{figure}

The result of the calibration of the amplitude of parametric pump can be also observed by measuring the cavity spectrum while sweeping the pump frequency $\omega_p$. Fig. S2 shows in-phase (I) and out-of-phase (Q) quadratures of the cavity spectrum near the resonant frequency of the cavity as a function of the parametric pump frequency, before and after the calibration. Before the calibration, in addition to the avoided crossing at 2.7 GHz due to the mode coupling between the cavity and R transmon, several wriggles appear in both I and Q indicating that the amplitude of the parametric pump is not uniform in the range of swept frequency. After the calibration, on the other hand, the wriggles disappear in both I and Q leaving the avoided crossing on top of a uniform background. This uniform background indeed indicates a constant amplitude of the pump in the range of swept frequency. It is worth mentioning that the calibration procedure described in this section can be performed by using the spectroscopy of either transmon and we obtained the same numerical factor $\lambda$ for each pump frequency regardless of our choice between two transmons.

\section{Measurements on the L transmon}

\begin{figure}[t]
\centering
\includegraphics[width=10cm]{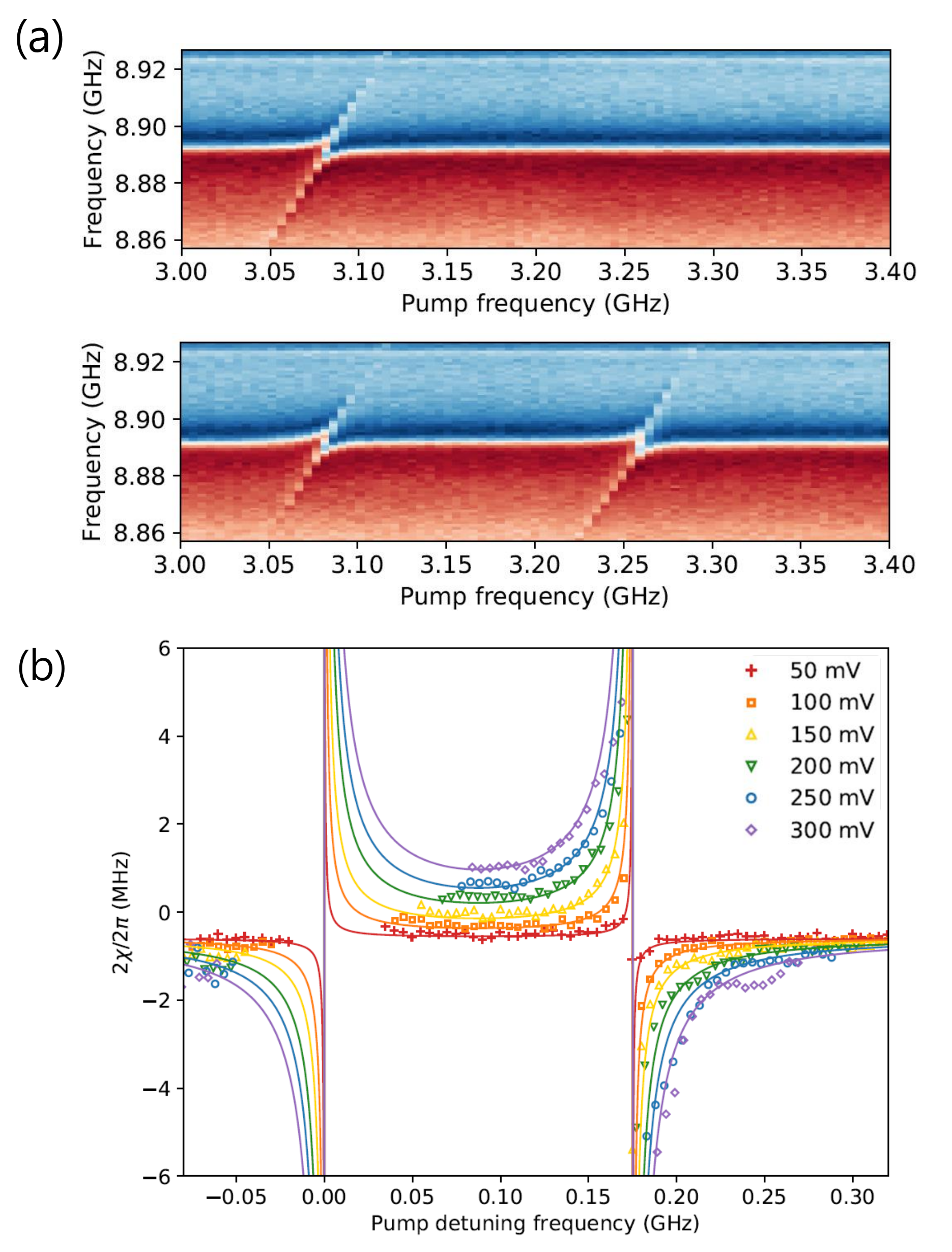}
\caption{(a) A typical cavity spectrum
while sweeping the pump frequency $\omega_p$. The upper and lower panel shows the result with the L transmon in the ground state $|g_L \rangle$ and the first excited state
$|e_L \rangle$, respectively. (b) The dispersive shifts as a function of the pump detuning frequency, $\Delta_{pL} = \omega_p - (\omega_C - \omega_L$), at various calibrated pump amplitudes: 50 (red), 100 (orange), 150 (yellow), 200 (green), 250 (blue), and 300 (purple) mV. The solid lines are the fits for the data from the theory. See the main text for the details.}
\end{figure}

We show the result of pulsed parametric dispersive readout of the R transmon in Fig. 2 in the main text. Here, we  show the result of the same set of measurements for the L transmon in Fig. S3. With the L transmon in the ground state $|g_L \rangle$, an avoided crossing appears with a coupling strength, $2g_{pL}/2\pi\approx 10$~MHz when the parametric pump frequency $\omega_p$ is near the difference frequency $\omega_p\approx|\Delta_L| = |\omega_C - \omega_L|$,  as shown in the upper panel of Fig. S3(a). After applying a $\pi$-pulse on the L transmon,  we can see in the lower panel of Fig. S3 (a) that another avoided crossing appears at a frequency centered about $\Delta_p= \omega_p - 
\Delta_L =-\alpha_L>0$. As the same for the R transmon, we repeated the same sets of measurement at various calibrated amplitudes, 50, 100, 150, 200, 250, and 300 mV. By comparing the two cavity spectra with the L transmon prepared at $|g_L \rangle$ and $|e_L \rangle$, we can extract the full dispersive shifts as shown in Fig. S3 (b), which shows a good agreement with the simple theory described in the main text.

\begin{figure}[t]
{\includegraphics[width=10cm]{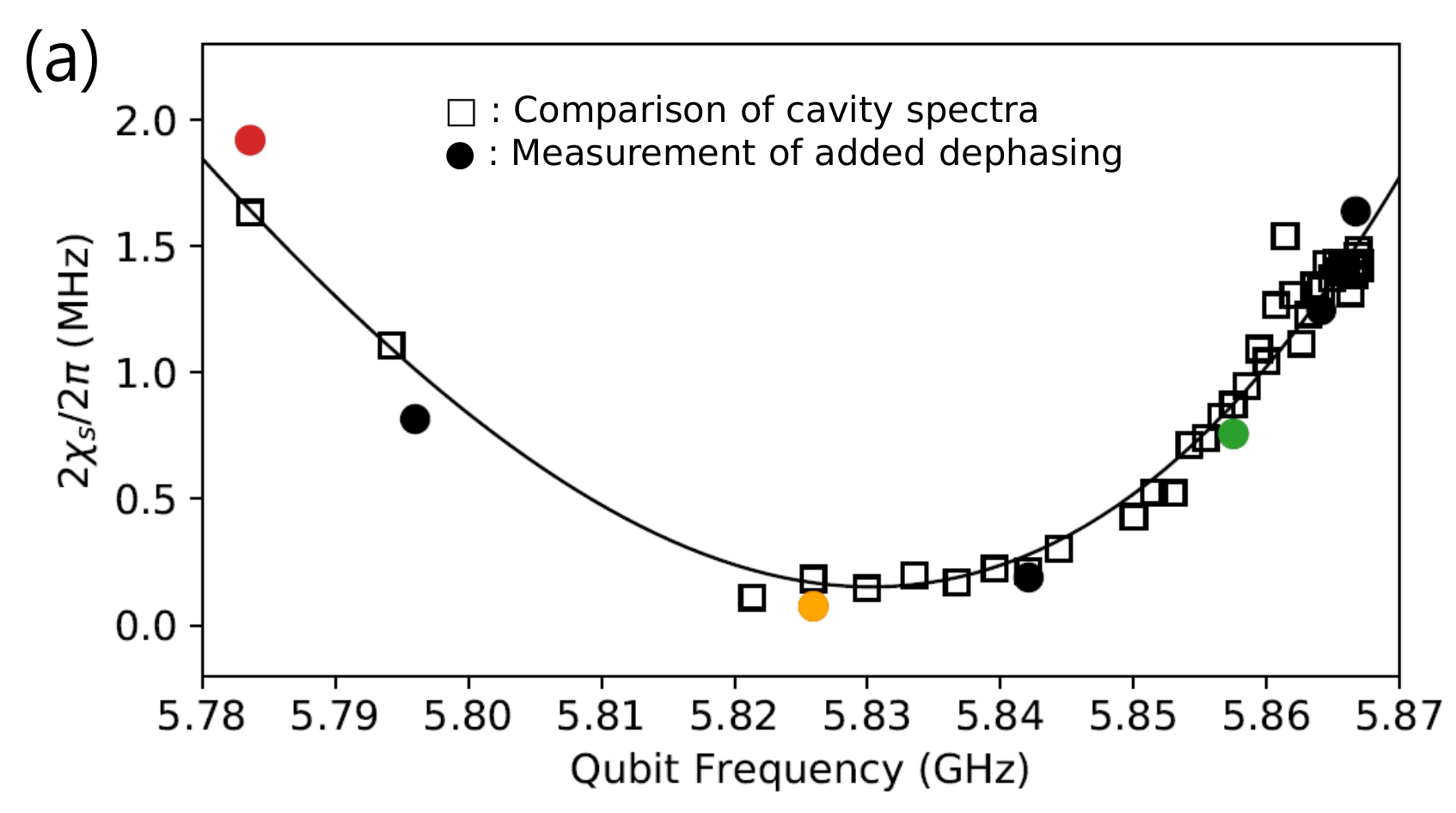}}
{\includegraphics[width=10cm]{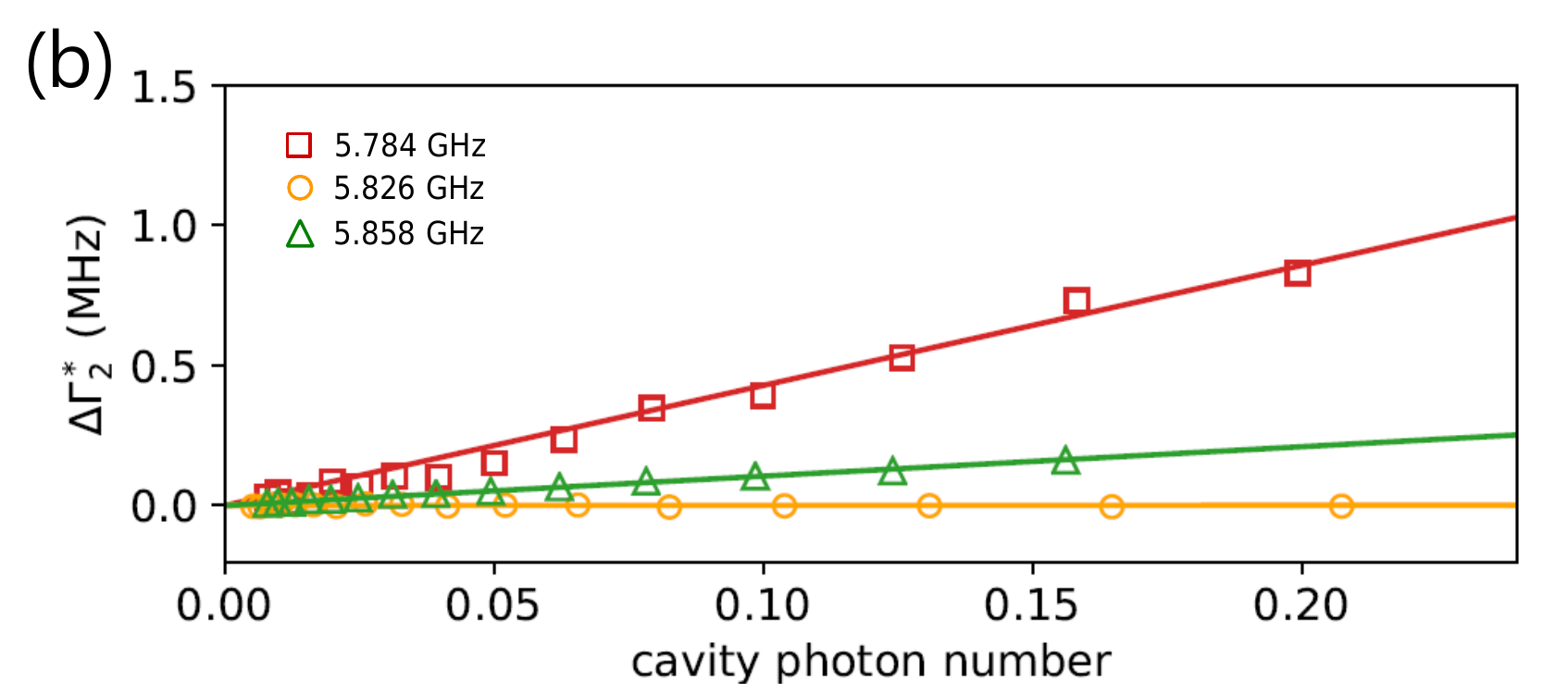}}
\caption{(a) The static dispersive shift $2 \chi_s/2 \pi$ as a function of the L transmon frequency. For empty squares, the dispersive shift was measured by comparing the cavity response with the L transmon in either state $|g\rangle _L$ or $|e\rangle _L$. For filled circles, the dispersive shift was extracted by measuring additional dephasing $\Gamma_{\phi L}$ while driving the cavity with a weak coherent state. (b) Additional dephasing as a function of average photon number $\bar{n}$ of a weak coherent drive at three different flux biases where the R tranmon frequencies are 5.784 (red square), 5.826 (yellow circle), and 5.858 GHz (green triangle), respectively. Solid lines are linear fit to the data. }
\end{figure}

In addition, we show the static dispersive shift $\chi_{s}$ as a function of the L transmon frequency in Fig. S4(a) . For the open squares, we directly measured the dispersive shift by comparing the cavity response with the L in either state $|g_L\rangle $ or $|e _L\rangle$. The solid line comes from predictions based on our circuit model. In Fig. S4(b), we measured additional dephasing $\Gamma_{\phi L}$ of the L transmon while driving the cavity with a weak coherent state with average photon number $\bar{n}$ for each flux bias, which shows linear dependences on $\bar{n}$ as expected.
Since the additional dephasing is given by the rate $\Gamma_n=8\bar{n}\kappa\chi_s^2/(\kappa^2 + 4\chi_s^2)$, where $\bar{n}$ is the average number of (coherent state) photons in the cavity, we can extract $\chi_{s}$ associated with the L transmon. This result is plotted as filled circles in Fig. S4(a) and agrees well with the direct measurements.  

\section{Parametric dispersive shifts at different flux biases}

We also performed parametric dispersive readout at different flux biases. As an example, Fig. S5(a) shows the dispersive shifts $\chi$ as a function of pump detuning frequency extracted by comparing the cavity spectrum with the R transmon at $|g_R \rangle$ and $|e_R \rangle$, at a flux $\phi= 0.27$. Each trace can be fitted to a curve $\chi = \chi_s + \chi_{pk}$ ($k=L$, $R$). Here, $\chi_{s}$ is static dispersive shift which is independent of pump frequency and $\chi_{pk}$ is parametric dispersive shift given by
\begin{equation}
    \chi_{pk} = \frac{g_{pk}^2}{\Delta_{pk}}\frac{\alpha_k}{(\alpha_k+\Delta_{pk})},
\label{eq_4}
\end{equation}
which is analogous to the static dispersive case with the identification $g_{sk}\rightarrow g_{pk}$ and $\Delta_k\rightarrow \Delta_{pk}$.~\cite{Koch07}. From the fit, we extracted $\chi_{s} \simeq$  -1.5 MHz which agrees well with the value obtained from a separate measurement. In addition, we also obtained the value of $g_{pR}$ at each pump amplitude. We repeated the measurements at two other flux biases, $\phi=$-0.37 and -0.27, and plotted the $g_{pR}$ as a function of the pump amplitude at all three flux biases, as shown in Fig. S5(b). Extracting the slope of $g_{pR}$ on the pump amplitude at different flux biases allows us to confirm the dependence of the relative magnitude of parametric coupling on the flux bias. The solid red line was obtained from the respective dependence of the R transmon frequency and the cavity frequency on the flux bias $\phi$ and by applying the relation $g_{pR} \propto \sqrt{(\frac{d\omega_R}{d\phi})(\frac{d\omega_C}{d\phi})}$.

\begin{figure}[t]
\centering
\includegraphics[width=16cm]{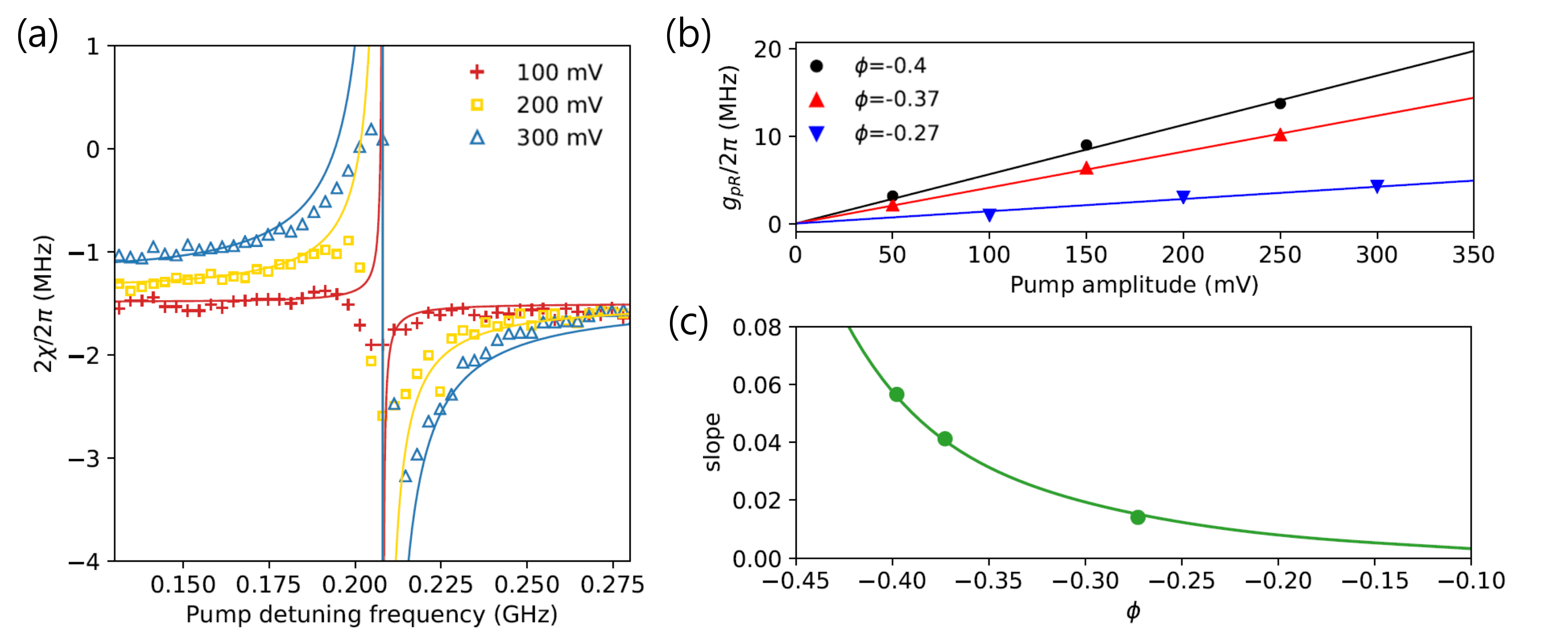}%
\caption{(a) The dispersive shifts as a function of the pump detuning frequency, $\Delta_{pR} = \omega_p - (\omega_C - \omega_R$), at various calibrated pump amplitudes: 100 (red), 200 (yellow), and 300 (blue) mV. The data was taken at $\phi = -0.27$. (b) Extracted $g_{pR}$ as a function of pump amplitude at three different flux biases: $\phi= -0.4$ (black), $-0.37$ (red), and $-0.27$ (blue). The solid lines are linear fits to the data. (c) The slope of each linear fit in (b) plotted as a function of the flux bias $\phi$. The red solid line is estimation based on the relation $g_{pR} \propto \sqrt{(d\omega_R/d\phi)(d\omega_C/d\phi)}$.  }
\end{figure}

\section{Parametric dispersive readout on higher level of transmon transition}

Parametric dispersive readout can be even extended to higher transmon levels, which leads to the modification of the Eq. (S3) to
\begin{equation}
    \chi_{pk} = \frac{g_{pk}^2}{\Delta_{pk}}\left[\frac{\alpha_k}{(\alpha_k+\Delta_{pk})}-\frac{\Delta_{pk}}{(2\alpha_k+\Delta_{pk})}-\frac{\Delta_{pk}}{(3\alpha_k+\Delta_{pk})}\cdots\right],
\label{eq_4}
\end{equation}
To demonstrate that, we compare the cavity spectrum measured while sweeping the pump frequency $\omega_p$ with the R transmon prepared in state $|e_R \rangle$ and $|f_R \rangle$. If we apply a $\pi$-pulse with frequency $\omega_{ge}$ of the R transmon prior to the measurement of the cavity spectrum, the R transmon is prepared in the first excited state $|e_R \rangle$. In this case, an avoided crossing appears at $\omega_p \simeq$ 2.86 GHz which corresponds to $\Delta_p  = - \alpha_R$, as presented in the upper panel of Fig. S6 as well as in Fig. 2(a) of the main text. On the other hand, if we apply a $\pi$-pulse with frequency $\omega_{gf}/2$ of the R transmon instead, the R transmon is prepared in the second excited state $|f_R \rangle$ by a two-photon absorption process. As shown in the lower panel of Fig. S6, we now observe an additional avoided crossing in the cavity spectrum at $\omega_p \simeq$ 3.04 GHz which corresponds to $\Delta_p  = - 2 \alpha_R$. This result indicates that the parametric dispersive readout can be performed on the higher levels of a transmon with a proper choice of the frequency and amplitude of the parametric pump, which can be extended as readout scheme for a system such as ``qudit".

\begin{figure}[t]
\centering
\includegraphics[width=10cm]{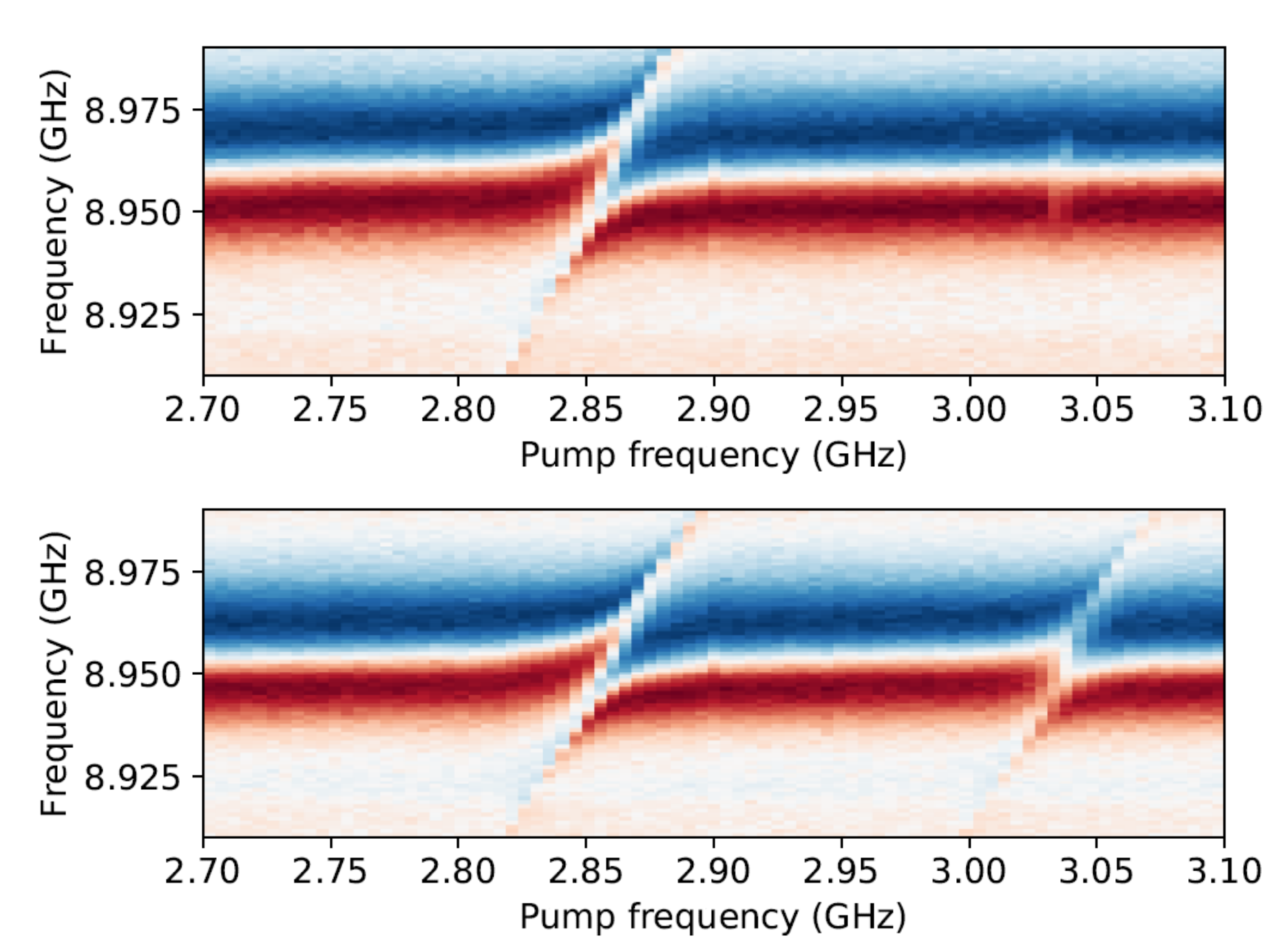}%
\caption{Cavity spectrum while sweeping the pump frequency $\omega_p$. The upper and lower panel shows the result with the R transmon in state $|e\rangle _R$ and $|f\rangle _R$ prepared by appyling a $\pi$-pulse with frequency $\omega_{ge}$ and $\omega_{gf}/2$ respectively, prior to the measurement of the spectrum. Data was taken at the same flux bias $\phi= \phi_c=$ -0.386 as the one shown Fig. 2(a) in the main text.}
\end{figure}